\documentclass{JINST}

\usepackage{bm}

\newcommand{\cov}{\mbox{$\mathrm{Cov}$}}
\newcommand{\Cov}{\mbox{$\mathrm{Cov}$}}

\title{Characterization and correction of\\charge-induced pixel shifts in DECam}

\author{Daniel Gruen$^{a,b}$\thanks{Corresponding author.}, Gary~M. Bernstein$^c$, Mike Jarvis$^c$, Barnaby Rowe$^d$, Vinu Vikram$^{c,e}$, Andr\'{e}s~A. Plazas$^{f,g}$ and Stella Seitz$^{a,b}$\\
\llap{$^a$}University Observatory Munich,\\
Scheinerstrasse 1, 81679 Munich, Germany\\
\llap{$^b$}Max Planck Institute for Extraterrestrial Physics,\\
Giessenbachstrasse, 85748 Garching, Germany\\
\llap{$^c$}Dept. of Physics and Astronomy, University of Pennsylvania,\\
Philadelphia, PA 19104, USA\\
\llap{$^d$}Dept. of Physics and Astronomy, University College London,\\
Gower Street, London, WC1E 6BT, UK\\
\llap{$^e$}Argonne National Laboratory,\\
9700 South Cass Avenue, Lemont, IL 60439, USA\\
\llap{$^f$}Brookhaven National Laboratory,\\
Bldg 510, Upton, NY 11973, USA\\
\llap{$^g$}Jet Propulsion Laboratory, California Institute of Technology,\\ Pasadena, CA 91109, USA\\
E-mail: \email{dgruen@usm.uni-muenchen.de}}

\abstract{Interaction of charges in CCDs with the already accumulated charge distribution causes both a flux dependence of the point-spread function (an increase of observed size with flux, also known as the brighter/fatter effect) and pixel-to-pixel correlations of the Poissonian noise in flat fields. We describe these effects in the Dark Energy Camera (DECam) with charge dependent shifts of effective pixel borders, i.e.~the Antilogus et al.~(2014) model, which we fit to measurements of flat-field Poissonian noise correlations. The latter fall off approximately as a power-law $r^{-2.5}$ with pixel separation $r$, are isotropic except for an asymmetry in the direct neighbors along rows and columns, are stable in time, and are weakly dependent on wavelength. They show variations from chip to chip at the 20\% level that correlate with the silicon resistivity. The charge shifts predicted by the model cause biased shape measurements, primarily due to their effect on bright stars, at levels exceeding weak lensing science requirements. We measure the flux dependence of star images and show that the effect can be mitigated by applying the reverse charge shifts at the pixel level during image processing. Differences in stellar size, however, remain significant due to residuals at larger distance from the centroid.}


\begin{document}

\section{Introduction}

An idealized telescope focal plane is covered by a grid of equal-sized pixels that have a linear response to the flux density at their corresponding position in the sky. Charge counts contain Poissonian noise that is independent between any two pixels. The images produced by such a camera then, up to noise, represent the sky convolved with a kernel. The latter, called point-spread function (PSF), can be measured from the observed profiles of point sources. The only difference between point sources, e.g. stars, of different flux levels is an amplitude by which the PSF profile is re-scaled.

Real CCDs with large well depth exhibit deviations from this picture, two of which have been known for some time. In exposures of homogeneous flux density (flat fields), the variance $V$ at moderate to high flux levels is not linear in the flat level $\mu$, unlike what is expected for a Poissonian process \cite{2006SPIE.6276E..09D}. Secondly, the size of star images has been found to increase with flux (known as the brighter/fatter effect), e.g. in science verification data from the Dark Energy Camera (DECam), but also in many other cameras (e.g.  MegaCam and LSST candidate sensors \cite{2014JInst...9C3048A}, Euclid candidate sensors \cite{2014arXiv1412.5382N} and the Wendelstein Wide Field Imager \cite{2014ExA...tmp...43K}).

The brighter/fatter effect is particularly problematic for ongoing and future large weak lensing programs such as the Dark Energy Survey (DES, \cite{2005astro.ph.10346T}), the Hyper Suprime-Cam Survey (HSC, \cite{2012SPIE.8446E..0ZM}), the Kilo Degree Survey (KiDS, \cite{2013Msngr.154...44J}), or the Large Synoptic Survey Telescope (LSST, \cite{2008arXiv0805.2366I}). The reason for this is that a systematic misrepresentation of the PSF in bright stars (that are used for PSF modeling) biases galaxy shape measurement, to a level that can exceed the acceptable systematics (cf. Section~3 and, e.g., \cite{2008MNRAS.391..228A}). Careful correction of this effect is therefore of great importance.

Two physical phenomena have been proposed and examined with analytical and numerical models \cite{2014JInst...9C3057H,2014SPIE.9150E..17R} that could be the cause for both observations: (1) lateral electric fields due to differences in electrical potential between pixels of different charge count (put more simply, mutual deflection of accumulating charges) and (2) an increase in lateral diffusion due to lower drift fields in pixels that are partially filled \cite{2014JInst...9C3057H}.

\cite{2014JInst...9C3048A,2015arXiv150101577G} (hereafter A14) have proposed a phenomenological model for charge self-interaction in CCDs that describes both the variance non-linearity and the brighter/fatter effect. In their model, accumulating charge shifts the effective pixel borders. In a flat field image, a pixel that has at any point during an exposure collected more charges than its neighbors due to positive contributions from Poissonian noise will, effectively, shrink. This decreases the variance and introduces a positive covariance with neighboring pixels. The charges already present in the image of a star shift pixel borders inward, such that the measured image on the pixel grid increases in size. Using laboratory and on-sky measurements with several different cameras, A14 have shown this model to reproduce the observed effects reasonably well. Both lateral electric fields and increased lateral diffusion due to lowered longitudinal electric fields are described by the same phenomenological model.

One of the instruments studied by A14 was DECam \cite{2006SPIE.6276E..08D,holland2007,2008SPIE.7014E..0ED,2008SPIE.7021E..07D,2010SPIE.7735E..1RE,2010SPIE.7735E..5CK,2012SPIE.8446E..11F,2012PhPro..37.1332D}, the 3 sq. deg. camera mounted at the prime focus of the Blanco 4m telescope at the Cerro Tololo Inter-American Observatory (CTIO), currently used for DES and community observing programs. The mosaic consists of 62 fully depleted 2K$\times$4K science CCDs with a thickness of 250$\mu$m. During the science verification phase of DECam, we detected a flux dependence of observed sizes of star images, in line with the A14 study, at a level relevant for DES science requirements.

In this work, we aim at correcting the effects of charge self-interaction in DECam images at a level tolerable for DES weak lensing science. In Section~2, we qualitatively characterize the brighter/fatter effect from DECam star images. Section~3 introduces the A14 model and explains our measurement of the model parameters from flat fields. In Section~4 we discuss the effect of charge self-interaction in DECam on galaxy shape measurements for weak lensing. Section~5 describes tests of our correction for charge self-interaction by reverse application of the model. We summarize our findings in Section~6.

\section{Brighter/fatter effect in DECam}
\label{sec:phenomenology}
In this section, we make a phenomenological description of the flux dependence of the PSF in DECam data. 

\begin{figure}
\centering
\includegraphics[width=0.48\textwidth]{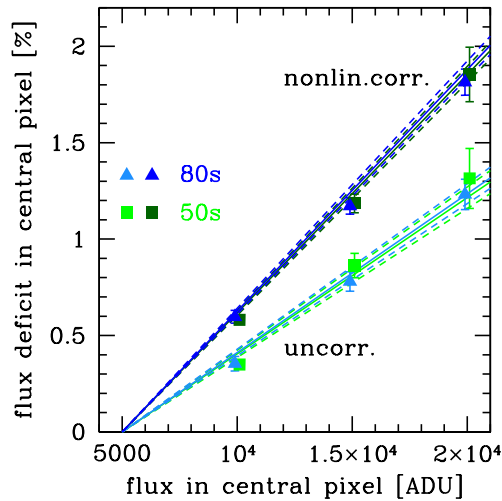}
\caption{Flux deficit of normalized PSF model in central pixel relative to a PSF model made with stars in peak flux (\texttt{FLUX\_MAX}) bin centered on 5000 ADU. We find that the flux deficit is approximately linear in flux level (solid lines show best linear fit, dashed lines 68 per cent confidence intervals), independent of exposure time (green squares: 50s, blue triangles: 80s), and increases when signal chain non-linearity is corrected for (dark green/blue: corrected, light green/blue: uncorrected).}
\label{fig:bf}
\end{figure}

We illustrate the effect in 2D by comparing the profiles of bright and faint stars in DECam images. The left panel of Fig.~\ref{fig:diffimg} shows the residuals of normalized profiles of stars at a peak flux (\textsc{SExtractor} \cite{1996A&AS..117..393B} parameter \texttt{FLUX\_MAX}) of 5000 and 20000, i.e. $(\mathrm{PSF}_{20000}-\mathrm{PSF}_{5000})/\mathrm{PSF}_{5000}$ where $\mathrm{PSF}_{N}$ is the PSF model made from stars around \texttt{FLUX\_MAX} $=N$. Flux is missing in the central region of the profile and re-appears on an annulus at $\approx2-4$~pix radius.

We use the flux deficit in the central pixel of the normalized PSF model as a metric to answer a number of questions. 
To this end, we use $i$ band images of the globular cluster $\omega$ Centauri taken with DECam during science verification on Feb 02 2013 with homogeneous seeing conditions. We process these as described in Section~\ref{sec:correction}, without the reverse charge shift model applied. Fig.~\ref{fig:bf} shows results as a function of \texttt{FLUX\_MAX} for star images with two different exposure times and for a data reduction scheme applying or not applying, respectively, a correction for the known signal chain non-linearity of DECam.\footnote{The non-linearity correction was determined by comparing the mean count levels of dome flats of different exposure time. At the 20k ADU level charge counts are approximately 1\% lower than expected from linear response.} We find that
\begin{itemize}
\item the relative flux deficit in the center of bright stars due to the brighter/fatter effect is well described as linear in flux, as indicated by the solid lines that are good fits to the data points;
\item the amplitude of the flux deficit does not depend on exposure time, but only on the measured count level, as visible from the agreement between the blue and green lines (dashed lines give confidence limits for the linear fit); this indicates that the charges must be moved while being collected, not while they are waiting for readout in the pixel potential well;
\item the effect is amplified when the high level signal chain non-linearity, a property of the readout electronics, is corrected for; relative photometry of stars in frames with different exposure times shows that correction for high level signal chain non-linearity is necessary for restoring the physical charge level present in the CCDs (cf. Bernstein, presentation at PACCD2013 conference\footnote{See \texttt{https://indico.bnl.gov/event/cosmo2013}.}); we will therefore apply the signal chain non-linearity correction in all following analyses.
\end{itemize}

\section{Model}
\label{sec:model}

We briefly introduce the model of A14 before we discuss our approach of constraining the model parameters in Sections~\ref{sec:flat} and \ref{sec:fit}.

The basic idea is that charges $q_{kl}$ in any pixel $(k,l)$ influence the path of newly generated charges on their way into the pixel well.\footnote{Note that this could be due to a combination of lateral fields and changes in lateral diffusion.} As a result, they induce a shift in the effective border of a pixel $(i,j)$, i.e., they change the mapping of physical area on the chip onto the pixels. 

\begin{figure}
\centering
\includegraphics[width=0.9\textwidth,bb = 0 129 561 332,clip]{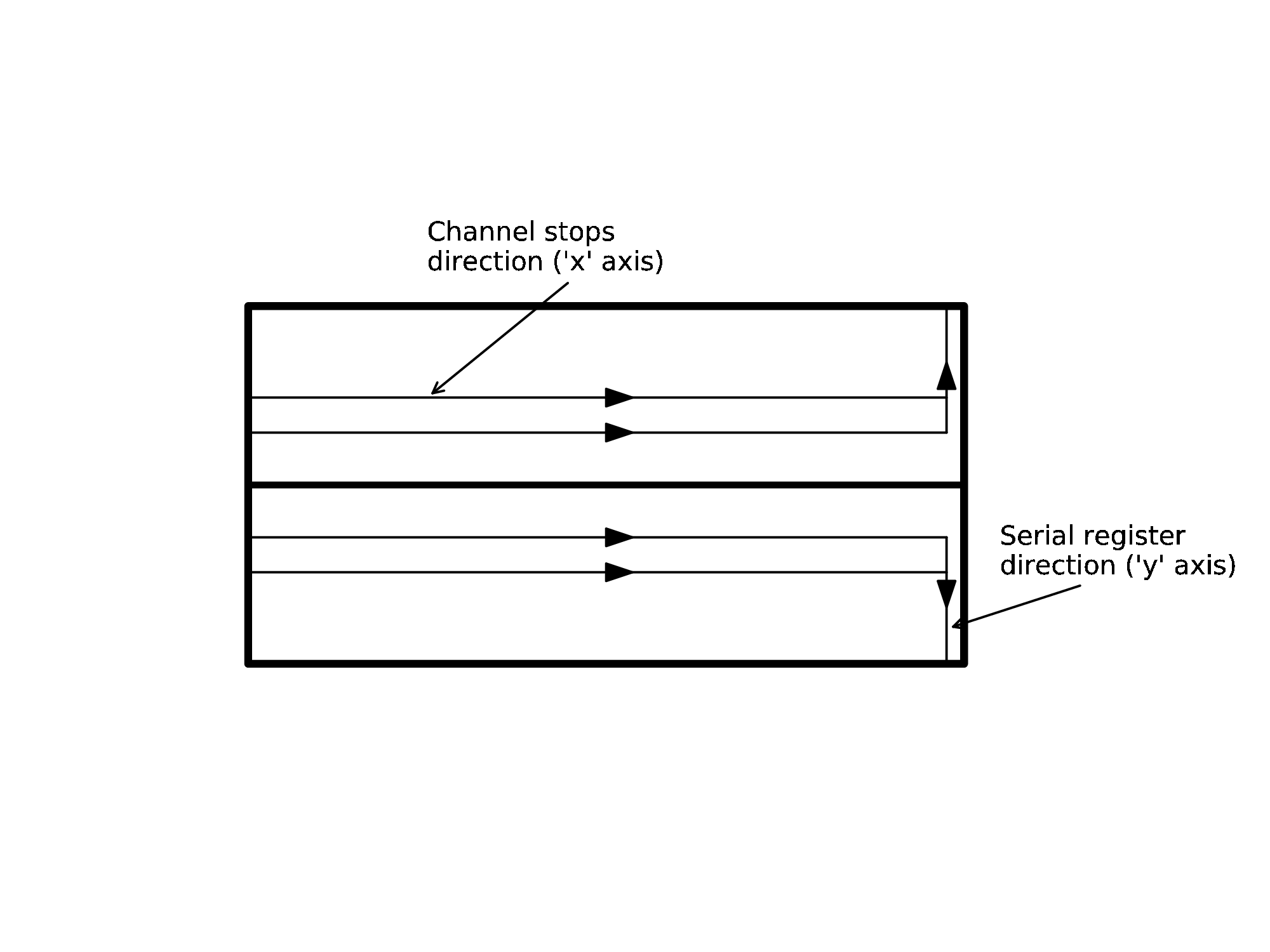}
\caption{Sketch of a DECam CCD and definition of our coordinate system, with the $x$ and $y$ axis parallel to the channel stops and the serial register, respectively. The split of the chip into two amplifier regions is indicated by the central horizontal line.}
\label{fig:chipsketch}
\end{figure}

Consider pixel $(i,j)$ and its four borders, which we label as $L=(-1,0)$, $R=(1,0)$ (left/right border to neighbors along $x$ axis) and $T=(0,1)$, $B=(0,-1)$ (top/bottom border to neighbors along $y$ axis). The shift in any of these we call $\delta_{ij}^{X}$, where $X\in\lbrace L,R,T,B\rbrace$. For a definition of our coordinate system, refer to Fig.~\ref{fig:chipsketch}.

We assume the border shift to be linear in the surrounding charge (including charge in pixel $(i,j)$ itself) and dependent on the \emph{lag} $(k-i,l-j)$ between pixel $(i,j)$ and $(k,l)$,
\begin{equation}
 \delta_{ij}^X=\sum_{kl}a^X_{k-i,l-j} q_{kl} \; .
\label{eqn:delta}
\end{equation}
The parameters of the model are the \emph{shift coefficients} $a^X_{ij}$. In all cases, $\delta_{ij}^X$ shall be defined such that the positive direction is outward from pixel $(i,j)$. 

The change in charges observed in pixel $(i,j)$ is the sum of $\delta_{ij}^R\frac{q_{ij}+q_{i+1,j}}{2}$ and the corresponding terms for $L,T,B$. Here, we have approximated the charge density at the pixel border by the mean of the neighboring pixels and defined the $\delta$ as fractions of a full pixel width. The observed charge $Q_{ij}$ then becomes
\begin{eqnarray}
Q_{ij}=q_{ij} \;\; &+& \;\; \frac{q_{ij}+q_{i+1,j}}{2}\sum_{kl}a^R_{k-i,l-j} q_{kl}
\;\; + \;\; \frac{q_{ij} + q_{i-1,j}}{2}\sum_{kl}a^L_{k-i,l-j} q_{kl} \nonumber \\ 
\;\; &+& \;\; \frac{q_{ij} + q_{i,j+1}}{2}\sum_{kl}a^T_{k-i,l-j} q_{kl}  
\;\; + \;\; \frac{q_{ij} + q_{i,j-1}}{2}\sum_{kl}a^B_{k-i,l-j} q_{kl} \; .
\label{eqn:Qq}
\end{eqnarray}

\subsection{A priori symmetries}

Assuming conservation of effective area, the area lost in one pixel is gained by its neighbor,
\begin{eqnarray}
\delta_{ij}^R=-\delta_{i+1,j}^L \; \nonumber \\
\delta_{ij}^T=-\delta_{i,j+1}^B \;
\end{eqnarray}

Eqn.~\ref{eqn:delta} fulfills these equalities for general charge distributions $q$ if and only if
\begin{eqnarray}
a_{ij}^L=-a_{i+1,j}^R \; , \nonumber \\
a_{ij}^B=-a_{i,j+1}^T \; .
\label{eqn:symmetry}
\end{eqnarray}

The number of free parameters is further reduced by two parity symmetries. 
The first is invariance of charge shifts under change of sign of one coordinate axis,
\begin{eqnarray}
a^{0,\pm1}_{i,j}=a^{0,\pm1}_{-i,j} \; , \nonumber \\
a^{\pm1,0}_{i,j}=a^{\pm1,0}_{i,-j} \; .
\end{eqnarray}
Another parity symmetry is that charges on opposite sides of the pixel border cause opposite shifts,
\begin{eqnarray}
a^{0,\pm1}_{i,j}=-a^{0,\pm1}_{i,\pm1-j} \; , \nonumber \\
a^{\pm1,0}_{i,j}=-a^{\pm1,0}_{\pm1-i,j} \; .
\end{eqnarray}
With these symmetries, the free parameters of the model are reduced to, e.g., $a_{ij}^R$ for $i>0$, $j\geq0$ and  $a_{ij}^T$ for $i\geq0$, $j>0$. Note that we discuss further empirically assumed symmetries, necessary to unambiguously constrain the parameters, in Section~\ref{sec:fit}.

\subsection{Flat field covariances}
\label{sec:flat}

The mixing of independent Poissonian processes $q$ into observed counts $Q$ in Eqn.~\ref{eqn:Qq} causes a correlation of noise in flat-field images. A14 collect the terms at first order in $a$ to find\footnote{Note that in their definition the $a$ are only half as large, origin of the factor of 2 here.}
\begin{equation}
\mathrm{Cov}(Q_{00},Q_{ij})=2\mu^2\sum_{X=T,B,L,R}a^X_{ij} \; ,
\label{eqn:flatcov}
\end{equation}
assuming $\mu$ is the mean charge count level and Poisson variance $[Var(q)=\mu]$ of charges in pixels in the flat image. We can thus use measurements of flat field covariances to constrain the $a$ parameters under suitable symmetry assumptions. Note that, in addition, there is a change in variance, which can be written using the above symmetries as
\begin{equation}
\Delta\mathrm{Var}(Q_{00})=-4\mu^2(a^R_{1,0}+a^T_{0,1}) \; ,
\label{eqn:dvar}
\end{equation}
equal with negative sign to the sum of all covariances introduced between $Q_{00}$ and the surrounding pixels.

Note that if we convert $Q$ and $\mu$ to quantities in ADU with some gain $g$, Eqns.~\ref{eqn:flatcov} and \ref{eqn:dvar} still hold with the same coefficients $a$. In Eqn.~\ref{eqn:Qq} we have to multiply the coefficients $a\rightarrow a'=ga$ to get the change in observed ADU.

\subsubsection*{Measurement pipeline}

We measure the covariances as follows, using all 10~s ($\approx15000$~ADU) dome flat fields in the r band from the full first year of DECam observations (2013-08-15 until 2014-02-09). We perform overscan and bias subtraction based on the nightly median bias. Inter-CCD cross-talk between the two amplifiers and the signal chain non-linearity are corrected for. Since cosmic rays are systematic contaminants of the observed covariances, we run \textsc{SExtractor} on each flat image to detect all groups of three or more pixels $3\sigma$ above the background. For each night, we mask the union of these $2\times2$ box-convolved object masks, a bad pixel mask, and edge distortions \cite{holland2009,2011ExA....29..135K,2014PASP..126..750P}. We re-scale each flat image so as to make median count levels of all frames taken in the same night match, which changes the levels below the per-cent level. From each pixel of the flat images, we subtract the mean of that pixel over all $n$ flats of the night.

From these residual images, we estimate covariances out to lags of $(8,8)$. We re-scale the residuals by $\sqrt{\frac{n}{n-1}}$ to correct the bias of the maximum likelihood ensemble covariance estimator. Since each chip has two readout channels, we make measurements separately for the 118 halves of chips in DECam,\footnote{Of the 62 chips, two (N30 and S30, i.e. \texttt{CCDNUM} 61 and 02) are damaged and one (S7, \texttt{CCDNUM} 31) has a time-varying offset, which is why we exclude them from our analysis.} i.e. on a chip by chip and amplifier by amplifier basis. 

Covariances are predicted by the A14 model to scale with the square of the count level $\mu$ (cf. Eqn.~\ref{eqn:flatcov}). We have verified this behavior of measured covariances from a PTC series of pairs of flat fields of different exposure times. For all following results, we therefore normalize our covariance measurements by $\mu^2$ to make them comparable despite somewhat varying flat levels between different nights and filters.

\begin{figure*}
\centering
\includegraphics[width=0.32\textwidth]{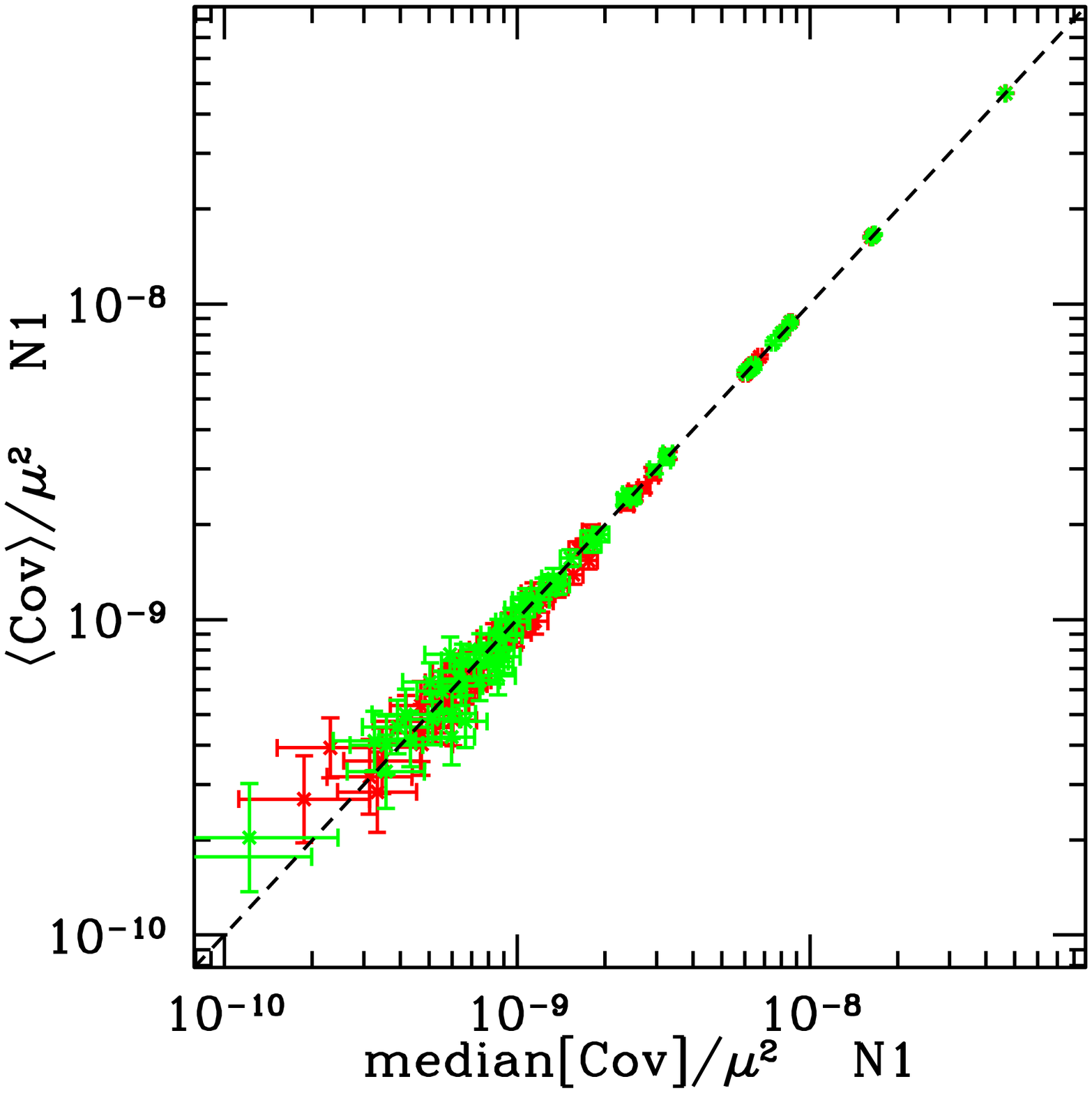}
\includegraphics[width=0.32\textwidth]{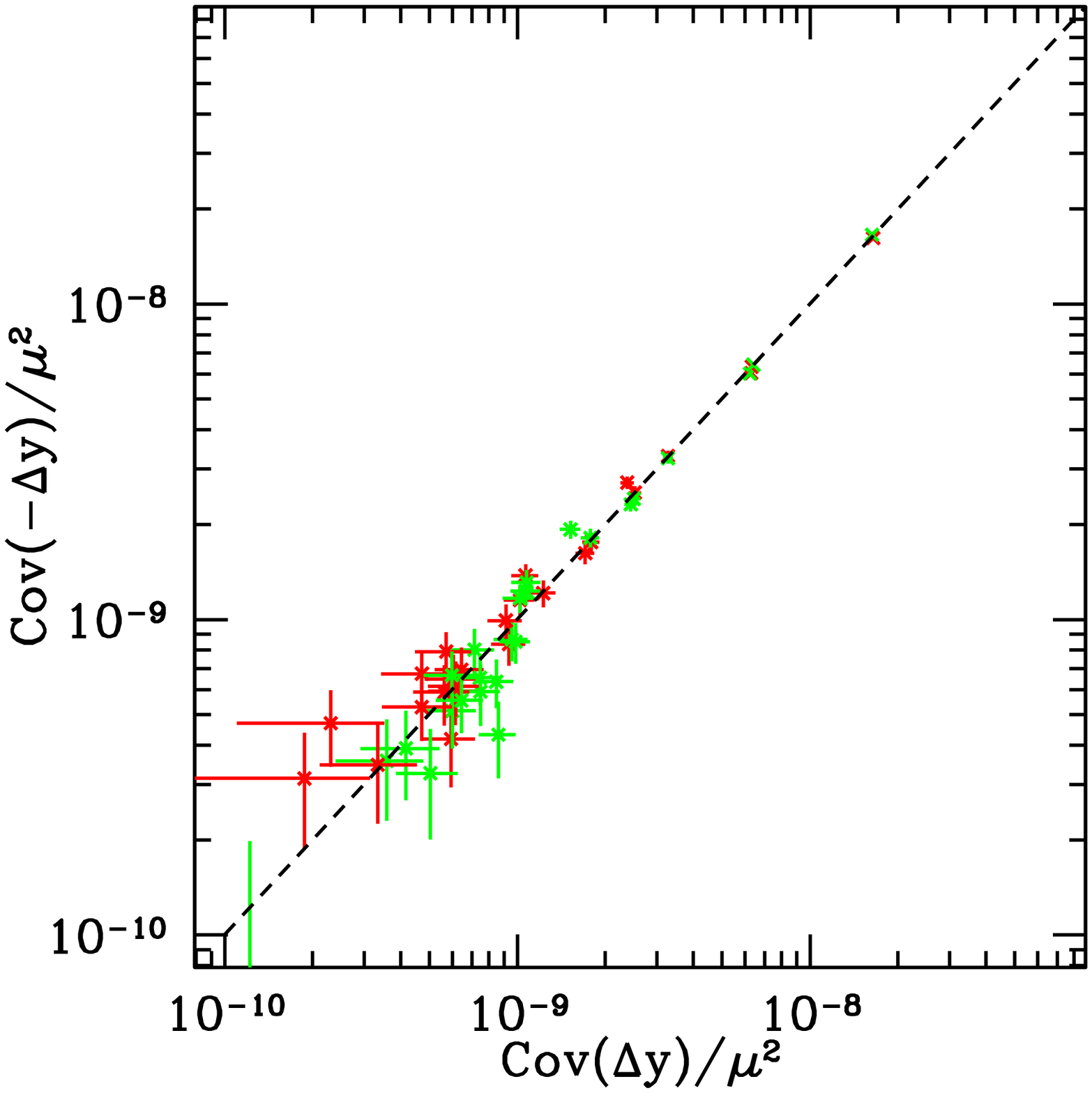}
\includegraphics[width=0.32\textwidth]{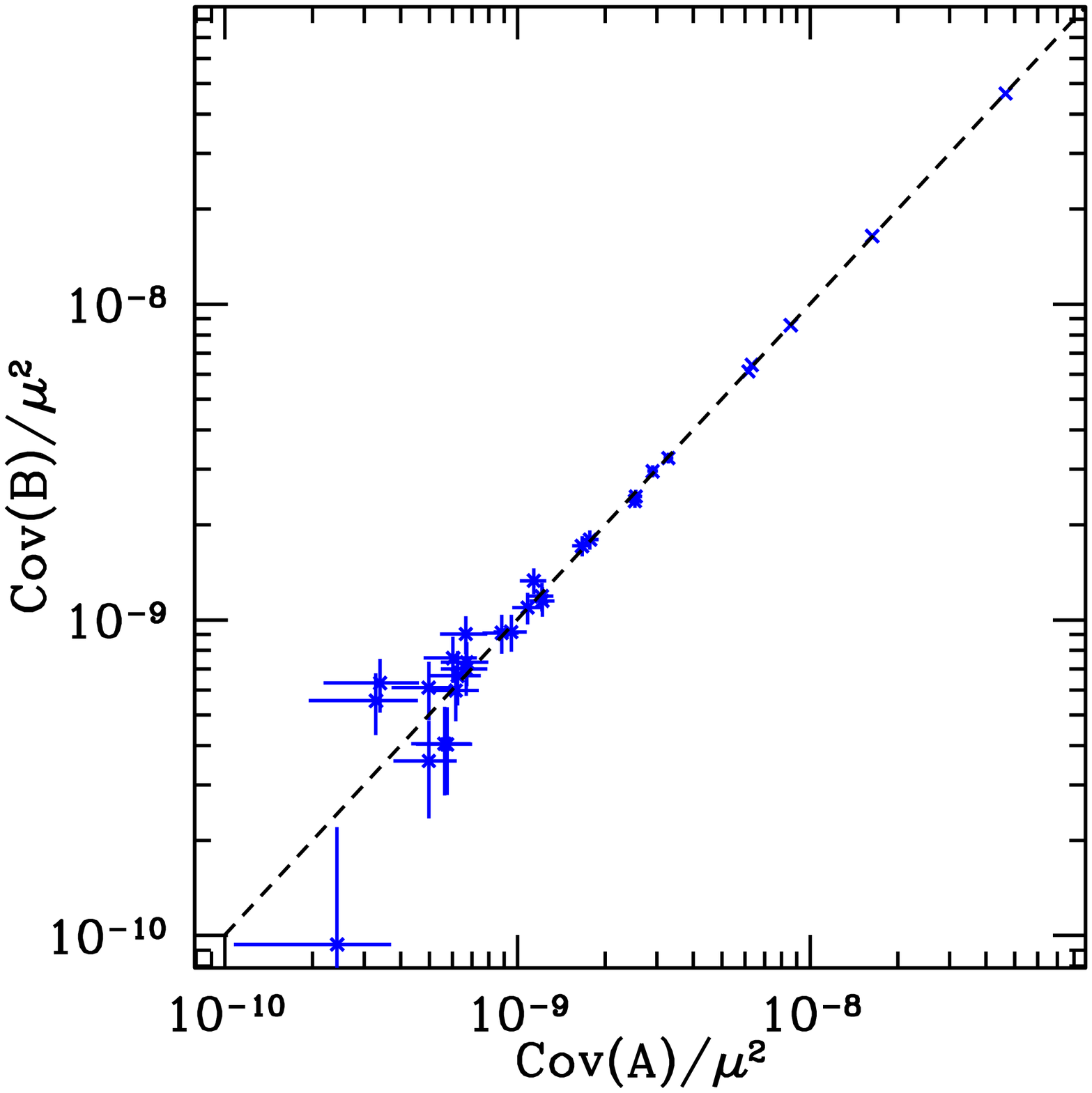}
\caption{Checks for systematics in measured covariances, shown for DECam chip N1. Left: median versus mean value of $\approx1200$ measurements for each lag (red: amplifier A, green: amplifier B). Center: covariances measured at lag $(i,j)$ versus $(i,-j)$. Right: covariances measured in amplifier A region versus amplifier B region. From these tests we conclude that the effect of outliers in measured covariances, artifacts, and differences within the chip are negligible.}
\label{fig:covsys}
\end{figure*}
\begin{figure*}
\centering
\includegraphics[width=0.48\textwidth]{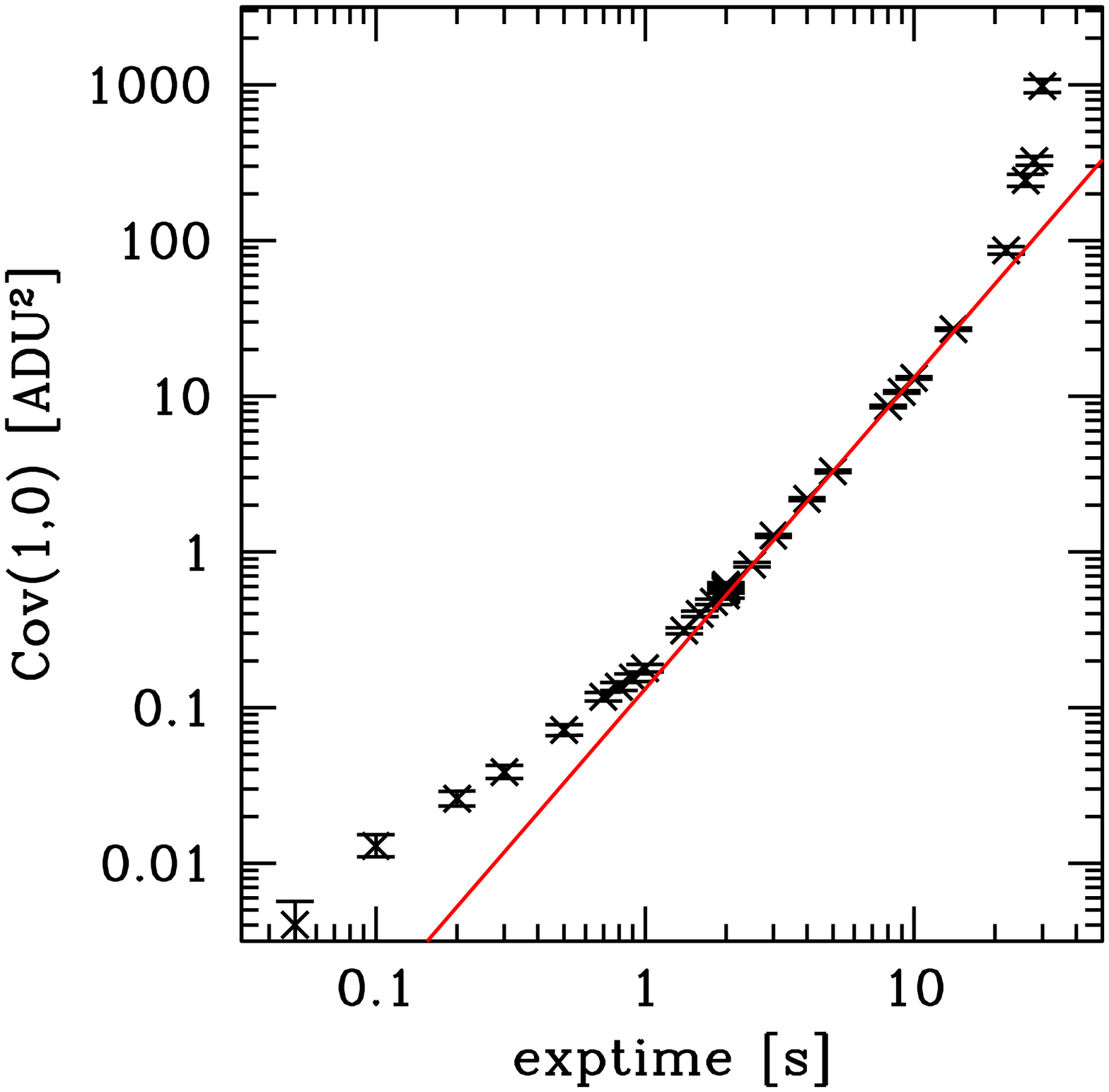}
\includegraphics[width=0.48\textwidth]{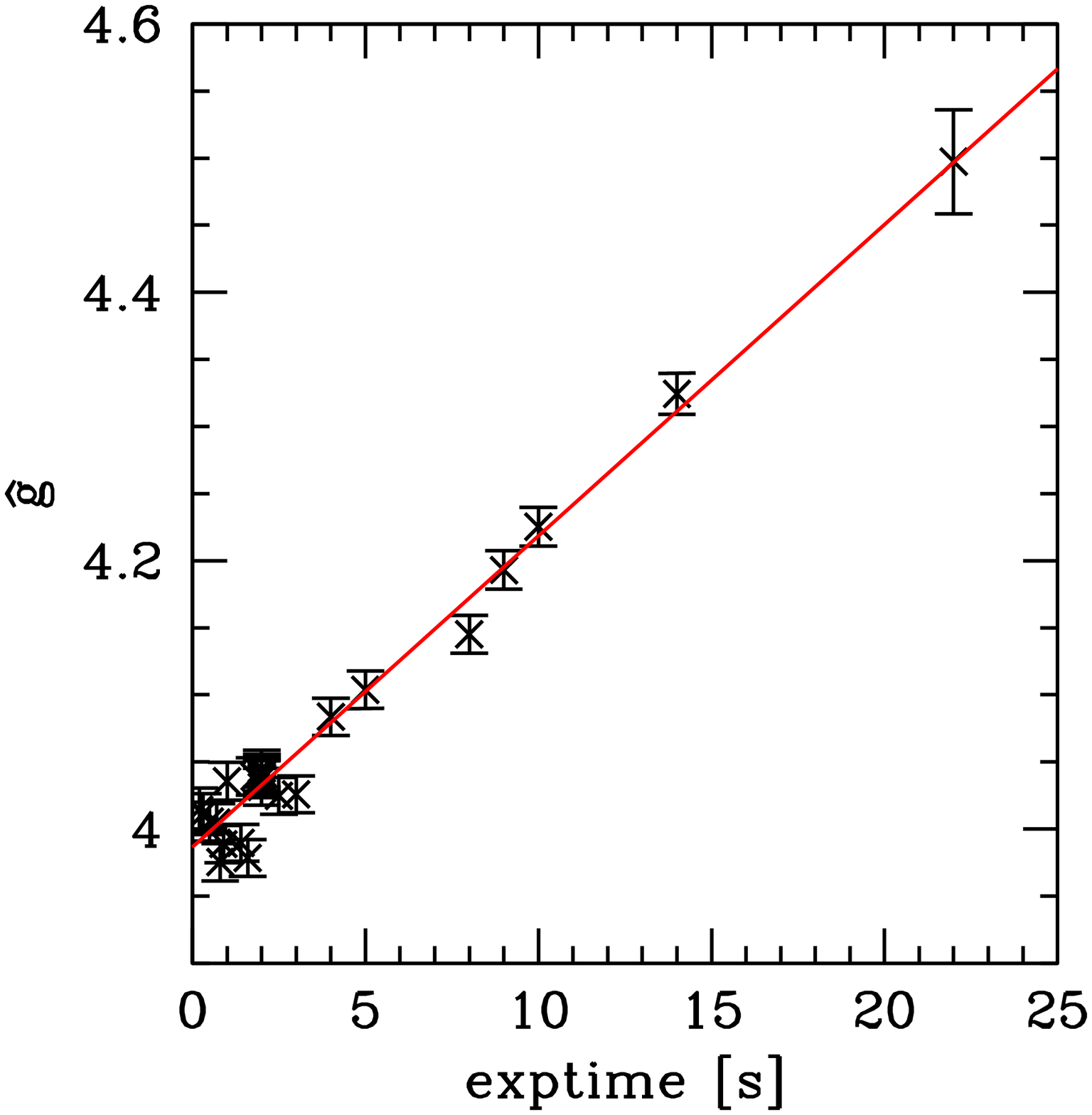}
\caption{Scaling of covariance and measured gain with count level, as measured in a series of pairs of flat field exposures with different exposure time. Left panel: covariance at (1,0) lag, with $\mu^2\propto t^2$ scaling around the $t=10$~s point indicated by red line. Right panel: measured gain $\hat{g}=\mu/V$, with the red line as a linear best fit to the data.}
\label{fig:fluxscaling}
\end{figure*}

\subsubsection*{Checks for systematics}

Each of the $\approx$1200 flat images yields an independent estimate, such that we can perform a number of systematic checks. For each chip, we make comparisons between different measurements that should give the same result:
\begin{itemize}
\item Since the statistics of our covariance estimator is very close to Gaussian, we expect no significant differences between using the mean and median of the $\approx1200$ independent measurements. Outliers, however, would influence the mean more strongly than the median.
\item Chirally-symmetric lags (e.g. $(1,1)$ and $(1,-1)$) measure covariances on disjunct sets of pairs of pixels that are, however, predicted to be the same from the model. 
\item If chips are homogeneous and the model is correct, then $\mathrm{Cov}(Q_{00},Q_{ij})/\mu^2$ should agree between the two halves of the chip read out by different amplifiers (cf. Fig.~\ref{fig:chipsketch}) despite their different gain values.
\end{itemize}
We find all these measurements to be consistent within the errors with systematic effects that are small compared to our targeted accuracy, out to large lags (cf. Fig.~\ref{fig:covsys}).

\subsubsection*{Flux scaling}

Covariances of the flat field Poisson noise are predicted to scale with $\mu^2$ according to the model of Eqn.~\ref{eqn:flatcov}. Measured gain, i.e.~the ratio of count level and variance, is predicted to rise linearly with flux due to the lowered variance. Other effects, such as correlated read noise, charge transfer inefficiency or other non-linearities in the signal chain, could cause different scalings. We test this by measuring covariances and gain at a range of count levels.

To this end, we analyze a photon transfer curve (PTC) series of pairs of flat field images with exposure times of $0.05\ldots30$s at a light level of $\approx 1500$ADU s$^{-1}$. Figure~\ref{fig:fluxscaling} shows covariances and gains measured from each pair, averaged over all valid chips. At very low levels, there is a floor of covariance below $\Cov<0.01$, which could be due to a small correlation of noise in the signal chain or a low-level non-linearity. At larger levels, correlations rise more quickly than predicted, potentially due to saturation and charge transfer effects. In the range of $2\ldots15$~s exposure time, however, the  $\Cov\propto\mu^2$ scaling describes the measurements well.

Likewise, measured gain is predicted by the model as linearly increasing with count level. Gain measurements in flats of $\approx$20~s exposure time overestimate the true gain by $\approx$10 \% and we correct the gains used in the DES data management system \cite{2012ApJ...757...83D,2012SPIE.8451E..0DM} for this effect.

\begin{figure}
\centering
\includegraphics[width=0.48\textwidth]{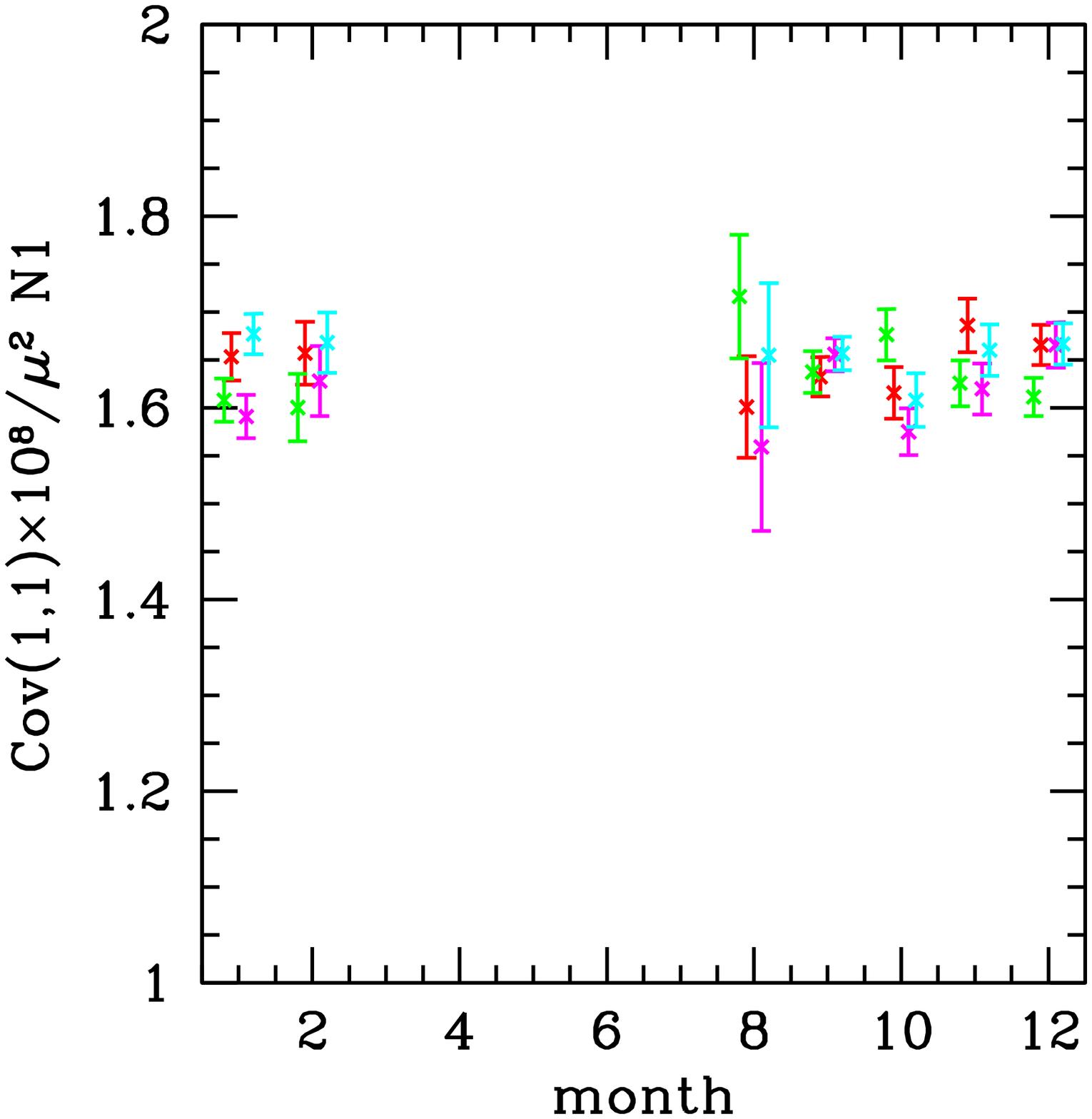}
\includegraphics[width=0.48\textwidth]{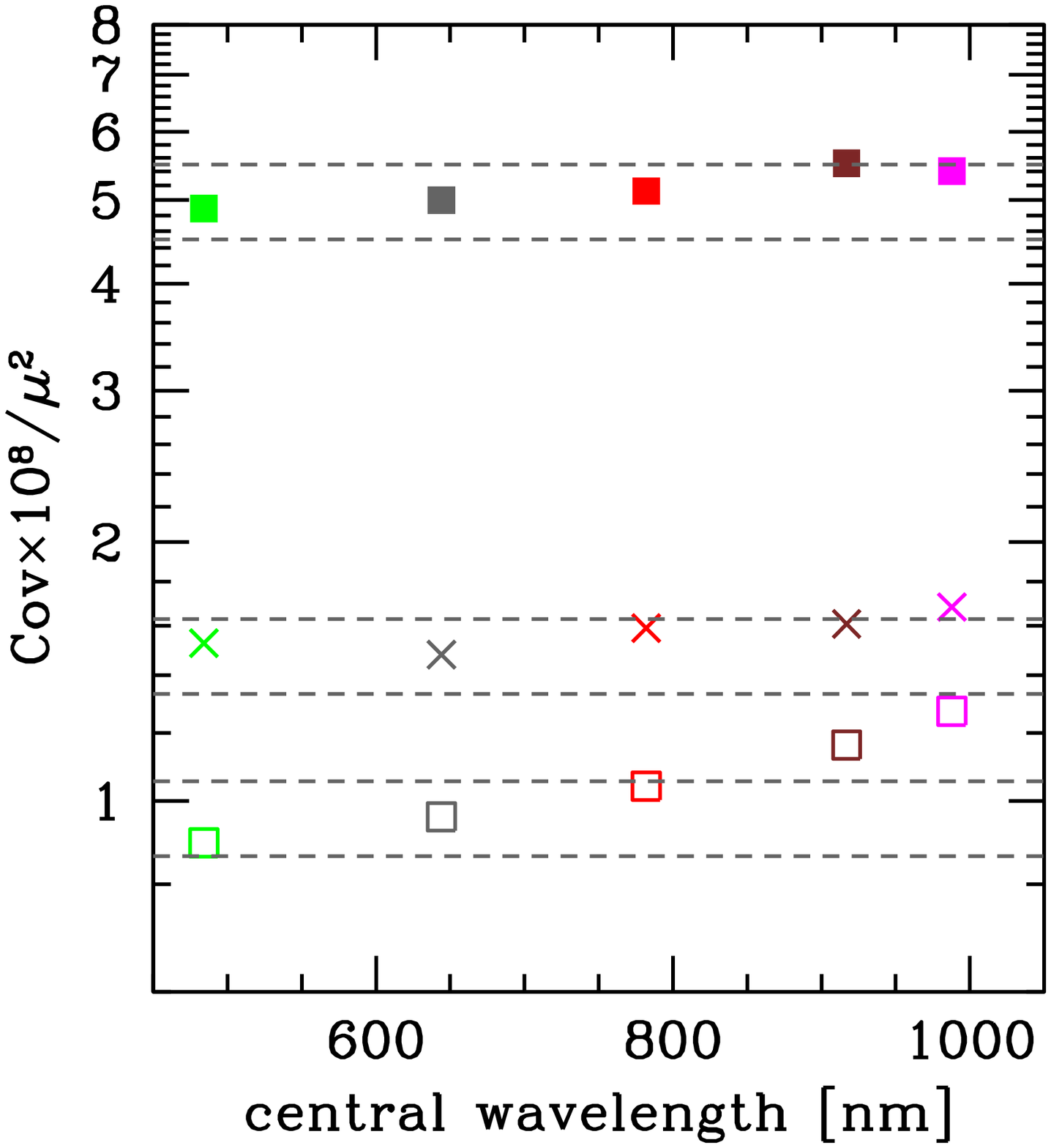}
\caption{Left: Test for time stability of flat field covariances for DECam chip N1. Shown are measurements in flat frames of each month of the DES season of the covariance at lag $(1,1)$ (red and green for amplifier A and B, respectively) and $(1,-1)$ (magenta and blue). The four measurements are offset on the time axis for readability. No significant time variability is observed. Right: Wavelength dependence of pixel-to-pixel covariances as measured in DECam g, r, i, z, and Y flat field images. Measurements are plotted at the respective central wavelength for lags $(1,0)$ (solid symbols), $(1,1)$ (crosses) and $(0,1)$ (open symbols). Intervals of $\pm10$\% around the r band measurements we use to fix the parameters of the model are indicated by dashed lines.}
\label{fig:covchromo}
\end{figure}

\subsubsection*{Wavelength dependence}

A14 found the noise correlations to be achromatic for the e2v-250 CCD run with nominal voltage configuration. This is in line with an effect that acts on the charges primarily in the last few $\mu$m of their drift path and therefore is almost independent of conversion depth. We measure covariances on g, i, z, and Y band flats from 20 nights to test for a potential wavelength dependence. Results for the three innermost lags, averaged over all 59 chips, are shown in the right panel of Fig.~\ref{fig:covchromo}. Both g and i measurements are consistent with the r band baseline model within 10\%. For the $(0,1)$ lag, the increase from g to Y band is at the 40\% level between the g and Y band. In $(1,0)$ and $(1,1)$ there are indications of an increasing trend, although at a weaker level.

\subsubsection*{Time stability}

We also test our measurements for time stability (cf. Fig~\ref{fig:covchromo}, left panel) by binning the measurements on a monthly basis. We find no evidence for time variation in any of the 59 chips used in our study.

\subsubsection*{Results}

\begin{figure}
\centering
\includegraphics[width=0.48\textwidth]{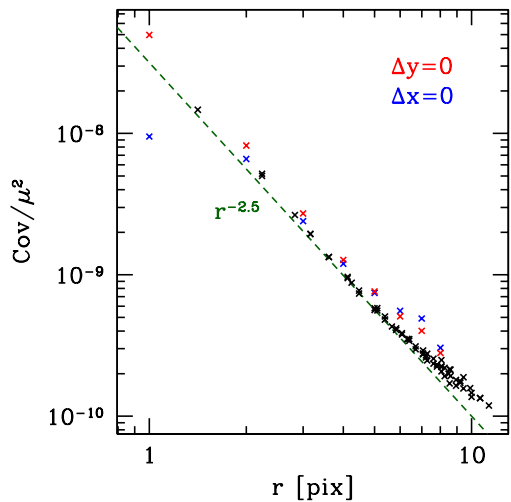}
\caption{Pixel-to-pixel covariance as function of distance $r=\sqrt{i^2+j^2}$. The long axis and readout direction is defined as $x$. Correlations with pixels along the same $x$ or $y$ coordinate are plotted in blue and red, respectively. A $r^{-2.5}$ power-law with arbitrary amplitude is indicated by the dashed green line.}
\label{fig:covavg}
\end{figure}

\begin{figure*}[t!]
\centering
\includegraphics[width=0.32\textwidth]{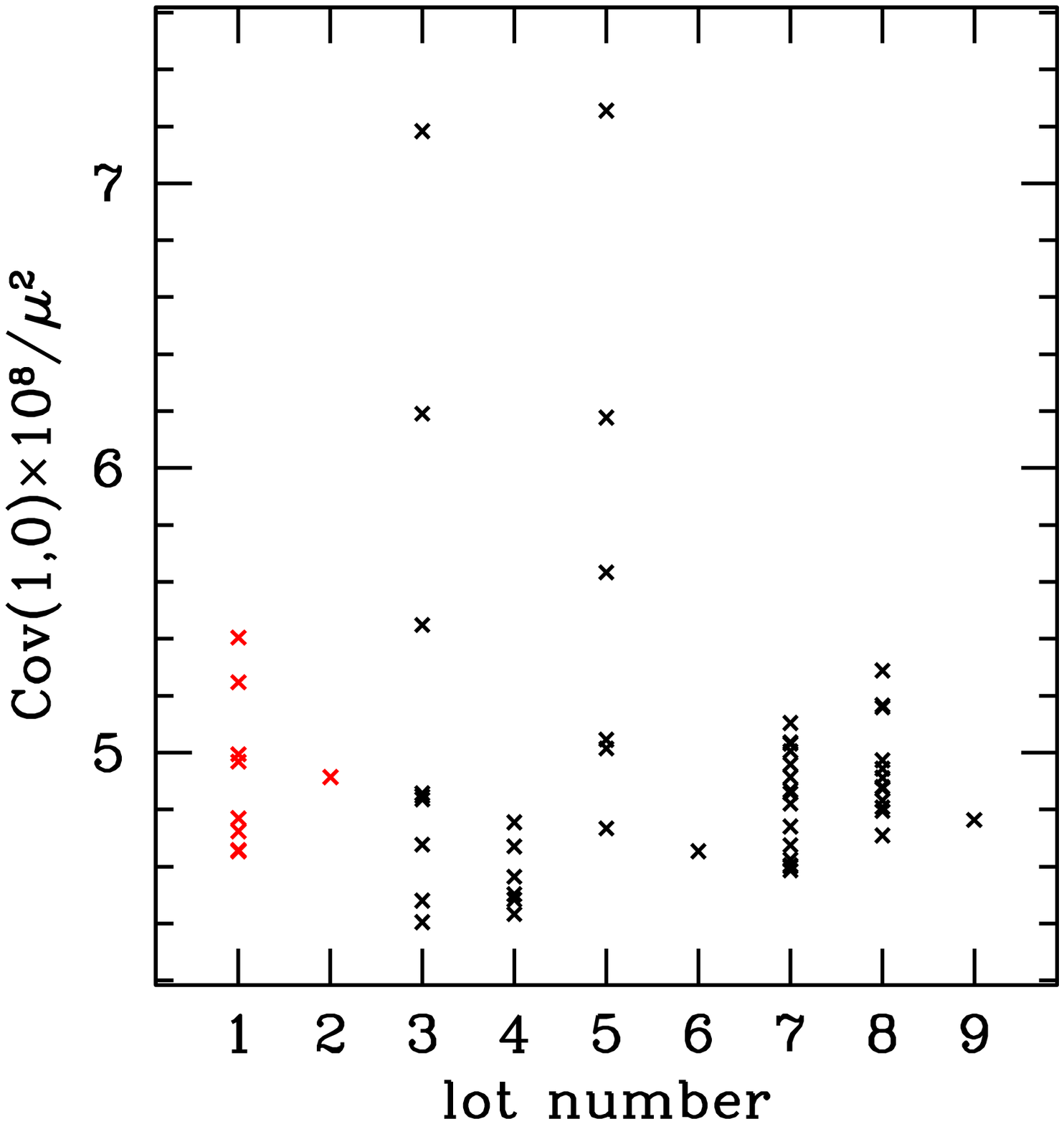}
\includegraphics[width=0.32\textwidth]{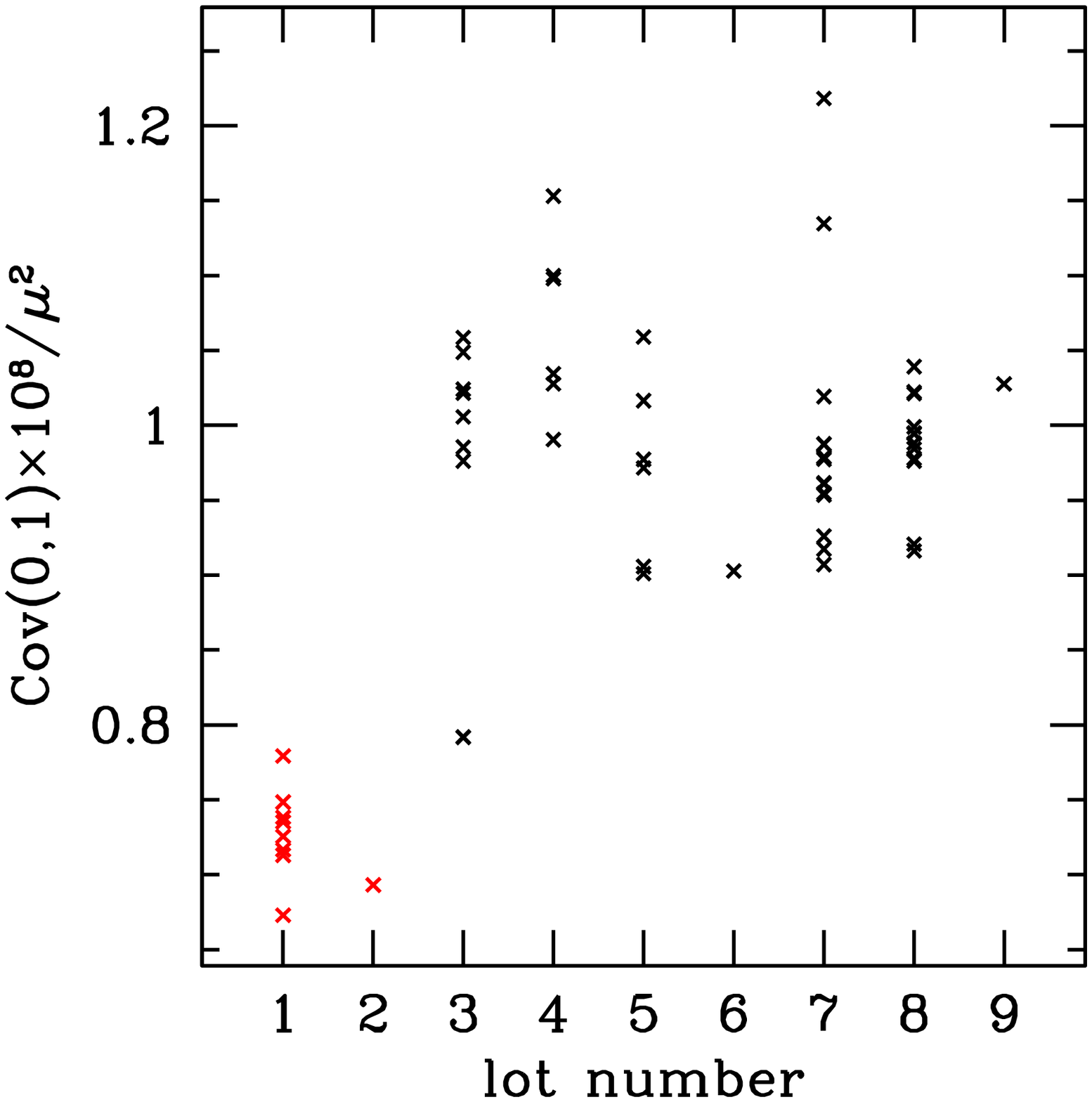}
\includegraphics[width=0.32\textwidth]{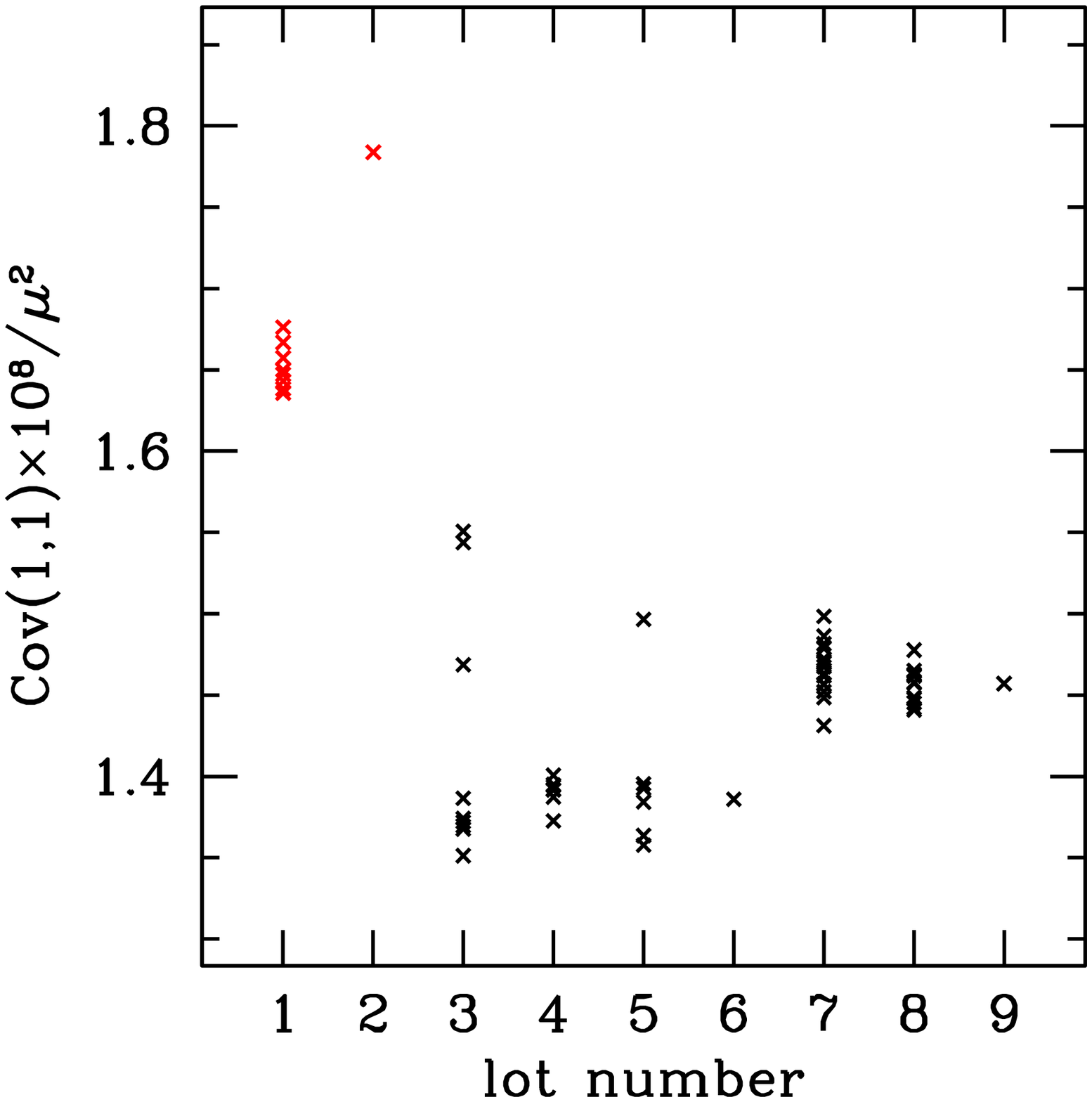}
\caption{$\mathrm{Cov}/\mu^2$ measured from flat fields for each of the DECam chips, plotted as a function to CCD production lot. Marked in red are the two high-resistivity lots. Left: $(1,0)$ lag (left/right neighbor along readout direction), center: $(0,1)$ lag (top/bottom neighbor across the channel stop), right: $(1,1)$ lag (diagonal neighbor).}
\label{fig:lots}
\end{figure*}

We show covariances as a function of lag distance $r=\sqrt{i^2+j^2}$, averaged over the 59 chips, in Fig.~\ref{fig:covavg}. We measure positive correlations with a signal-to-noise ratio of $\approx 15$ even at the outermost point $r=\sqrt{128}$. Off-axis covariances show a smooth power-law drop-off with radius, with indications for a break at around $r\approx5$. The behavior of on-axis coefficients is slightly different, with a higher amplitude and a low outlier at the $(0,1)$ lag, i.e. for neighbors across the channel stop. We note that the ratio $\Cov(1,0)/\Cov(0,1)\approx5$ is in the range the \cite{2014SPIE.9150E..17R} simulations predict for reasonable channel stop barrier dipole moments.

We test for variations of the level of covariances between different chips. Since all chips in DECam use the same nominal bias and clock voltages, the reason for such differences would have to be connected to chip properties. The science chips used in DECam were produced in nine different production lots. For simplicity of notation, we label the two high resistivity (10.6-14.2 k$\Omega$ cm) lots 107419 and 112094 as 1 and 2 and the remaining lots (with lower resistivity of 4.3-6.2 k$\Omega$ cm) 123194, 123195, 124750, 124753, 135959, 135960 and 135961 as 3-9 (in this order). Fig.~\ref{fig:lots} shows results for the three nearest lags. We find variations in amplitude at the $\approx20$\% level from lot to lot. Variations within the lots are significantly smaller, although with outliers, particularly at the $(1,0)$ and $(1,1)$ lag in lots 3 and 5. 

The two high resistivity lots have covariances $\approx20$\% higher (lower) than the low resistivity ones in the $(1,1)$ (the $0,1$) lag. Larger lags show a similar resistivity dependence as $(1,1)$. For the $(1,0)$ lag, no strong resistivity dependence is observed. This is evidence for a mixing of two effects with different dependence on resistivity and distance. While this suggests that resistivity is a relevant factor, other unknown lot-to-lot variations that coincide with resistivity by chance cannot be excluded, particularly since the high resistivity lots were cut from a different boule than the remaining CCDs.

\subsection{Assumed model symmetries}
\label{sec:fit}

The parameters of the model are the shift coefficients $a$ from Eqn.~\ref{eqn:delta}. They are connected to the measured covariances (see Section~\ref{sec:flat}) by Eqn.~\ref{eqn:flatcov}. With only the symmetries of Eqns.~\ref{eqn:symmetry}ff, however, the shift coefficients are not unambiguously constrained by flat field covariances (cf.~A14).

This is most comprehensibly exemplified by the following degeneracy. Finding all covariance terms that contain any $a^X_{ij}$ with $i<2$ and $j<2$
\begin{eqnarray}
\Delta\cov(Q_{00},Q_{00})    &=&-4\mu^2(a^R_{1,0}+a^T_{0,1}) \nonumber \\
\cov(Q_{00},Q_{0,\pm1})&=&2\mu^2(a^T_{0,1}-2a^R_{1,1}) + \ldots \nonumber \\
\cov(Q_{00},Q_{\pm1,0})&=&2\mu^2(a^R_{1,0}-2a^T_{1,1}) + \ldots \nonumber \\
\cov(Q_{00},Q_{1,1})   &=&2\mu^2(a^R_{1,1}+a^T_{1,1}) + \ldots \; 
\end{eqnarray}
we see that all covariances (including the variance defect $\Delta\cov$) are degenerate under the transformation
\begin{eqnarray}
a^R_{1,0}&\rightarrow& a^R_{1,0}+\Delta \nonumber \\
a^T_{0,1}&\rightarrow& a^T_{0,1}-\Delta \nonumber \\
a^R_{1,1}&\rightarrow& a^R_{1,1}-\Delta/2 \nonumber \\
a^T_{1,1}&\rightarrow& a^T_{1,1}+\Delta/2 \; 
\label{eqn:degeneracy}
\end{eqnarray}
with an arbitrary constant $\Delta$, which compensate the smaller shifts of one pixel border by larger shifts of the perpendicular ones. Clearly, however, the effect on inhomogeneous surface brightness images is not invariant under the above change on model coefficients.

Our approach is to assume two additional symmetries (the rotational and projection symmetry described below) that are suggested by the covariance measurements. For the above degeneracy, this entails setting $a^R_{1,1}=a^T_{1,1}$ and modelling the observed asymmetry in $\cov(Q_{00},Q_{0,\pm1})\neq\cov(Q_{00},Q_{\pm1,0})$ by differences in $a^R_{1,0}$ and $a^T_{0,1}$ only.

\subsubsection*{Rotational symmetry}

All off-axis lags rotated by $\pi/2$ have covariances measured to be equal at the level of agreement between the A and B sides of chips, i.e. $\cov(Q_{00},Q_{ij})=\cov(Q_{00},Q_{ji})$ for $i>0$, $j>0$. This symmetry is predicted by the model if 
\begin{equation}
a^R_{ij}=a^T_{ji} \; \; \forall i>0,\,j>0 \; .
\end{equation}
If we assume this symmetry, we forfeit almost half of our observables (because the covariances at lag $(i,j)$ and $(j,i)$ are now two measurements on the same combination of shift coefficients). In order to break the degeneracies, it would suffice to set $a^R_{i,i}=c_i a^T_{i,i}$, $i>0$, with a set of assumed ratios $c_i$. Due to the agreement of measured covariances at rotated lags, however, this would just yield a statistically consistent (yet noisier) model.

\subsubsection*{Projection symmetry}

The lateral electric field due to two charges at the same distance from a border should result from an electrostatic force that has the same amplitude but a different direction.\footnote{Note that the situation is different for lateral diffusion, which is affected by the change in the longitudinal electric field. For distant enough pairs, however, lateral electric fields appear to be the dominant effect.} The most similar pairs of charges in terms of distance are ones at lag $(i,j)$ and $(j,i)$. If we assume that the border displacement is proportional to the component of the force projected onto the normal vector of a border, we have (for $i>1$ or $j>1$)
\begin{equation}
a^R_{j,i}=r(i,j)a^R_{i,j} \; ,
\label{eqn:aRr}
\end{equation}
where
\begin{equation}
r(i,j) = \frac{j-0.5}{i-0.5} \sqrt{\frac{(i-0.5)^2+j^2}{(j-0.5)^2+i^2}} \; .
\label{eqn:r}
\end{equation}
For $a^T$, we write the analogous
\begin{equation}
a^T_{j,i}=r^{-1}(i,j)a^T_{i,j} \; .
\label{eqn:aTr}
\end{equation}

As we find from a first iteration of fitting, the fall-off of shift coefficients with radius is close to a power-law with slope $\alpha=-2$. We use that to refine Eqn.~\ref{eqn:r} as
\begin{equation}
r'(i,j) = r(i,j) \left(\sqrt{\frac{(j-0.5)^2+i^2}{(i-0.5)^2+j^2}}\right)^{\alpha} =\frac{j-0.5}{i-0.5} \left(\frac{(i-0.5)^2+j^2}{(j-0.5)^2+i^2}\right)^{3/2}\; 
\end{equation}
and insert $r'$ instead of $r$ in Eqns.~\ref{eqn:aRr} and \ref{eqn:aTr} for our final model.

\subsection{Fitting of coefficients}
\label{sec:fitting}

Assuming these symmetries, we have reduced the vector $\bm{a}$ of independent coefficients to $a^R_{1,0}$, $a^T_{0,1}$, $a^R_{1,1}$ and $a^R_{i,j}$ for $1<i\geq j\geq0$ (a total number of $\sum_{i=0}^n(i+1)=\frac{1}{2}(n+1)(n+2)$ when going out to a maximum of $n$ pixels distance). This matches the dimension of the vector $\bm{c}$ of independent measured covariances at the same lags.

Out to lags of $\Delta=3$ we use the direct measurement of covariances of each chip individually. At larger distances, the chip-wise signal-to-noise ratio of the covariances is low. We therefore assume a power-law fall-off of covariances with radius, as observed in Section~\ref{sec:flat},
\begin{equation}
\Cov(i,j)=A(i^2+j^2)^{-\beta/2}
\label{eqn:covpl}
\end{equation}
with $\beta=2.5$. We fit three independent amplitudes (one for lags along the $x$ and $y$ axis and one for off-axis pixels) to the covariances measured between 3 and 5 pixels separation. Covariances out to a maximum lag of $25$ are extrapolated from this model. 

Using Eqn.~\ref{eqn:flatcov} and the symmetries we express $\bm{c}$ as a linear function of $\bm{a}$, $\bm{c}=M\bm{a}$. We invert to find $\bm{a}=M^{-1}\bm{c}$ as a function of the measured and extrapolated vector of covariances.

We note that this approach is somewhat different from the one proposed by \cite{2015arXiv150101577G}, who model the radial fall-off of the shift coefficients in LSST candidate sensors by an exponential integral.

\subsubsection*{Tests}

We perform two consistency tests of the fitted model. 

\begin{figure}
\centering
\includegraphics[width=0.48\textwidth]{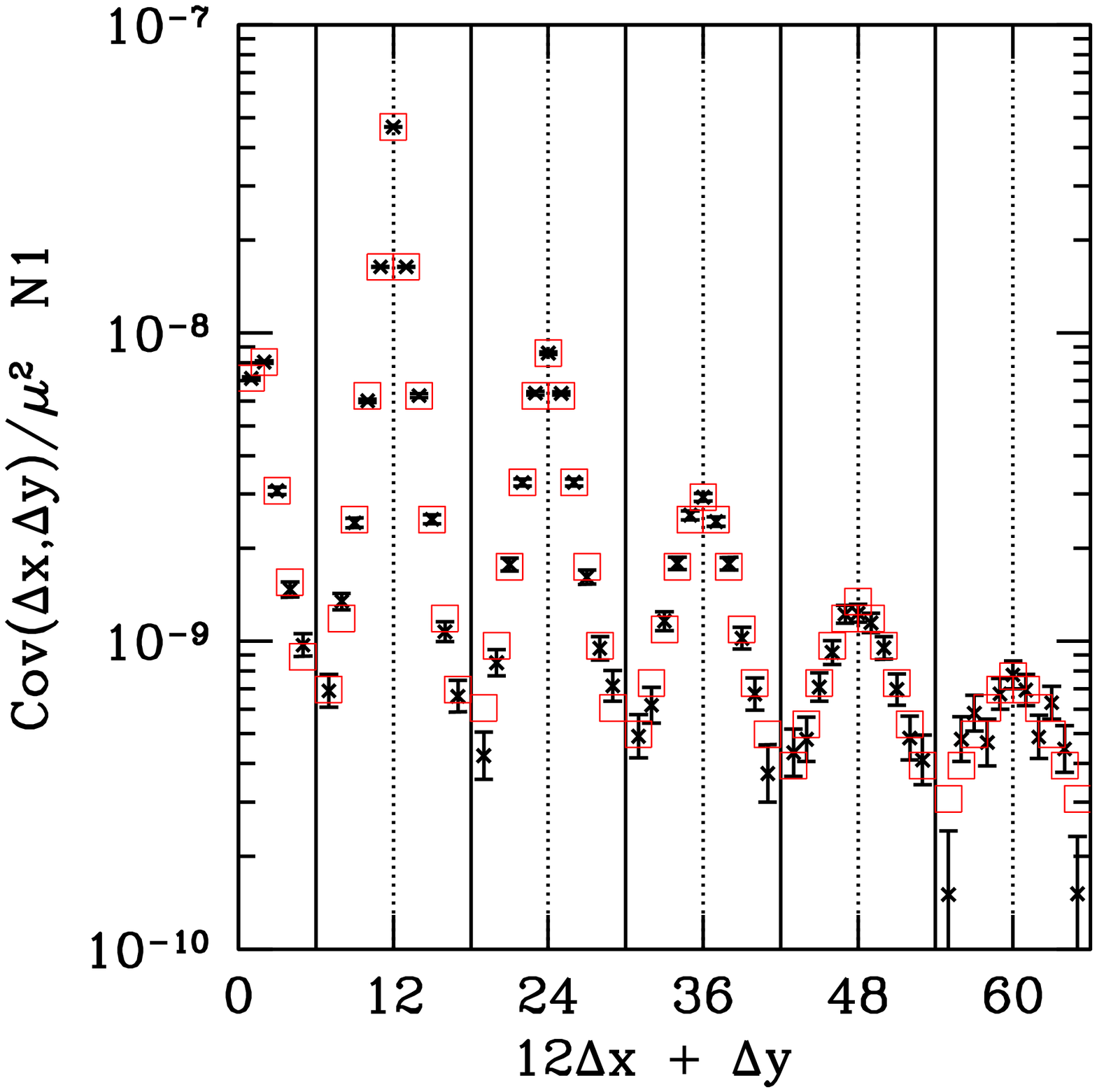}
\includegraphics[width=0.48\textwidth]{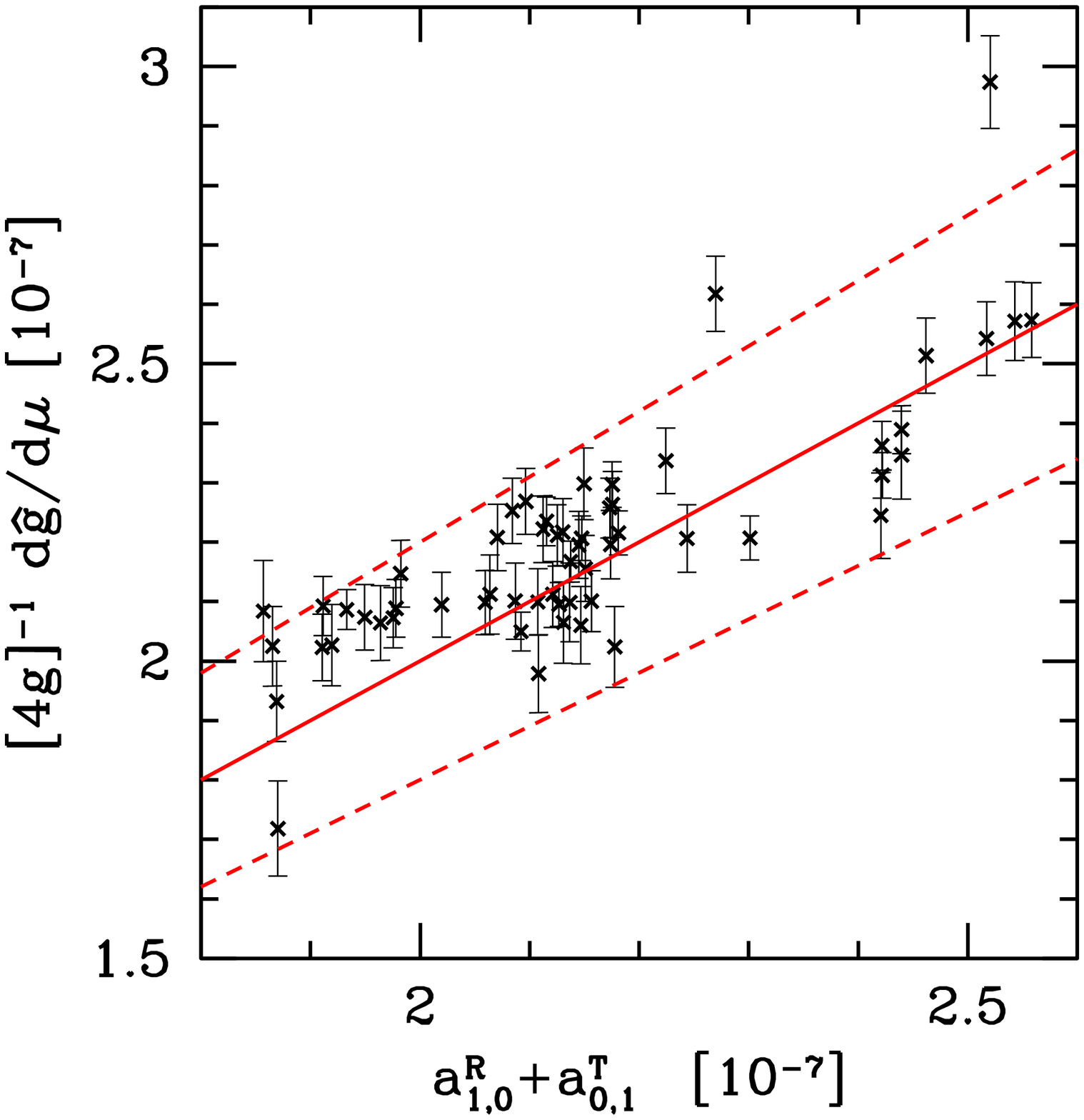}
\caption{Tests of the fitted shift coefficient model. Left panel: measured (black points with error bars) and model (red open squares) covariance coefficients for lags out to $5$ for CCD N1. Right panel: comparison of slope of observed gain $\hat{g}$ w.r.t. flat count level $\mu$ to sum of model coefficients for all chips. According to Eqn.~3.8, we expect the two quantities to relate with a factor of $4g$ (solid red line). The mean agreement is better than at the ten per-cent level (dashed red lines).}
\label{fig:dga}
\end{figure}

By comparing to measurements (Fig.~\ref{fig:dga}, left panel), we confirm that the model correctly reproduces (at lags $\leq3$) and extrapolates (at larger lags) the observed covariances. 

We further compare the fitted coefficients to our gain measurements from the PTC series (cf.~Fig.~\ref{fig:fluxscaling}). The deviation of $\hat{g}$ from the true gain $g$ is predicted to be linear in flux with a slope of $4g(a^R_{1,0}+a^{T}_{0,1})$. We show the measured slope from the PTC series plotted against model parameters for all chips in Fig.~\ref{fig:dga} (right panel), finding the two to be in agreement at the 10\% level. Note that the fitted $a^R_{1,0}+a^{T}_{0,1}$ are quite sensitive to changes in the shift coefficients at large radii, i.e. the maximum lag used or the power-law coefficient of Eqn.~\ref{eqn:covpl}.

\subsubsection*{Results}

\begin{figure*}
\includegraphics[width=0.32\textwidth]{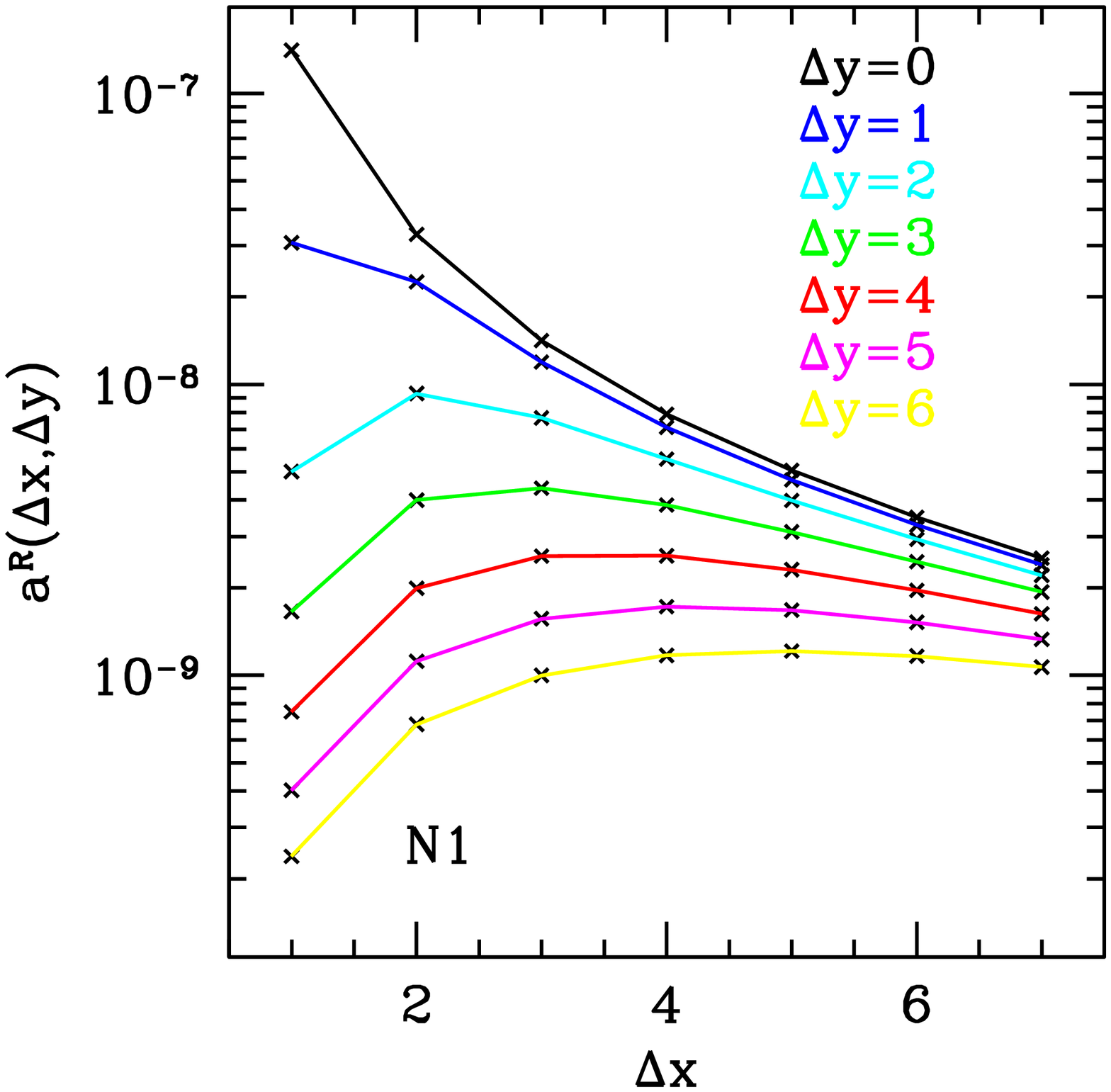}
\includegraphics[width=0.32\textwidth]{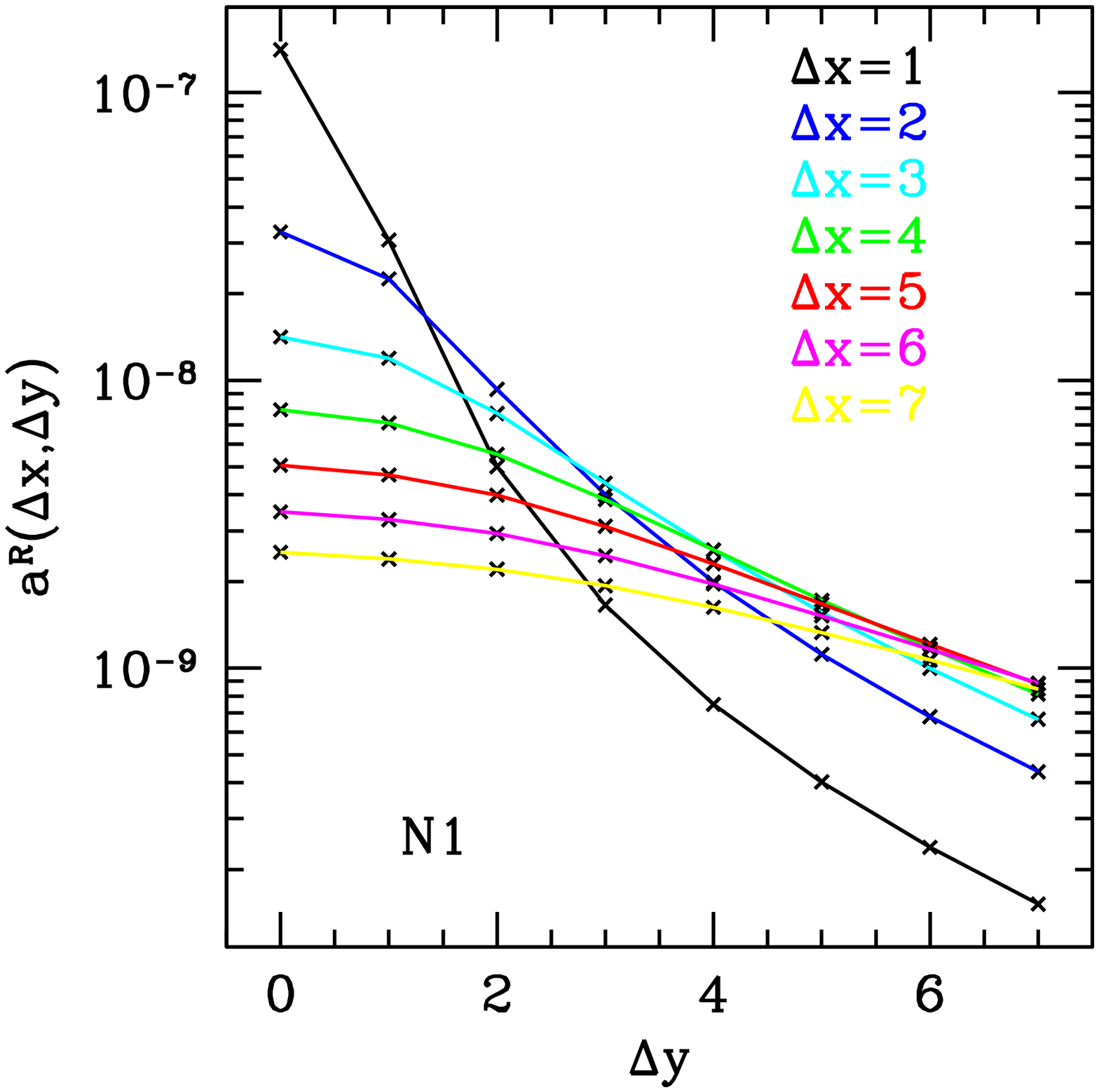}
\includegraphics[width=0.32\textwidth]{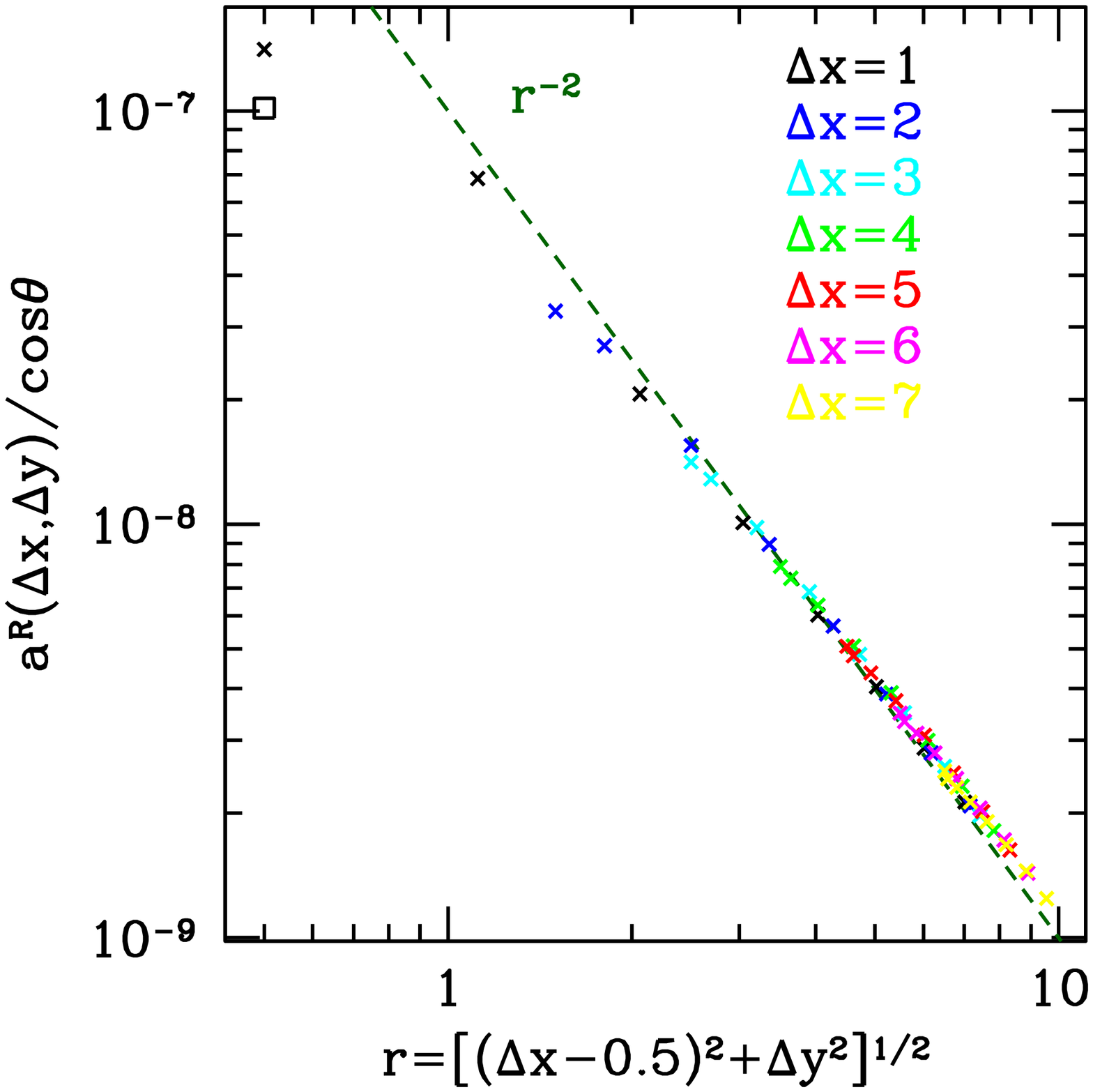}
\caption{Shift coefficients $a^R$ of chip N1 plotted as function of $\Delta x$ (left panel) and $\Delta y$ (center). Projection effects are corrected in the right panel, where we normalize coefficients by the cosine of the angle between pixel border normal vector and direction of the lag. This leads to a well-behaved $r^{-2}$-like power-law behavior (indicated by dotted green line) of the coefficients, with the neighbors $a^R_{1,0}$ and $a^T_{0,1}$ (the latter is overplotted with the open symbol) as outliers.}
\label{fig:axy}
\end{figure*}

We show the fitted shift coefficients $a^R$ of one exemplary chip in Fig.~\ref{fig:axy}. The fact that for rows at $\Delta y\geq 2$ there is a maximum at $\Delta x > 1$ (or, equivalently, that curves for constant $\Delta x$ cross as a function of $\Delta y$) is due to projection effects: at larger $\Delta x$ 
the separation increases, but the electrostatic force is aligned closer to the normal of the pixel border. Using the projection angle $\theta$, we correct for this effect by normalizing with $\cos\theta=[\Delta x-0.5]/[\sqrt{(\Delta x-0.5)^2+\Delta y^2}]$.

The de-projected coefficients $a^R_{ij}/\cos\theta$ appear to follow a single, power-law radius dependence (Fig.~\ref{fig:axy}, right panel). The coefficient $a^T_{0,1}$ is the strongest outlier from this trend. For chip N1, $a^T_{0,1}\approx1.0\times10^{-7}$ whereas $a^R_{1,0}\approx1.4\times10^{-7}$, indicating that an additional physical mechanism acts on directly neighboring pixels.\footnote{Compare also to the observed inverted resistivity dependence of the $(0,1)$ lag. Note, however, that from the PTC flat series we find all covariances to be consistent with a $\mu^2$ scaling as predicted by the charge shift model.} The moderate asymmetry of charge shifting is amplified by the differential nature of how the shift coefficients contribute to the pixel-to-pixel covariance to yield the strong asymmetry between $\cov(1,0)$ and $\cov(0,1)$ that we observed in Fig.~\ref{fig:covavg}.

\section{Effects on galaxy shape measurement}
\label{sec:sim}

One of the primary science applications of DES is to probe cosmic shear and the weak gravitational lensing effect of galaxies and clusters of galaxies. The measurement of gravitational shear $\gamma=\gamma_1+i\gamma_2$ is based on a measurement of galaxy shapes $\epsilon=\epsilon_1+i\epsilon_2$. The latter can be defined based on second-order moments of the pre-seeing surface brightness profile $I(\bm{\theta})$ of a galaxy centered at $\bm{\theta}=(0,0)$,
\begin{equation}
Q_{ij}=\frac{\int I(\bm{\theta})\theta_i\theta_j \mathrm{d}^2\theta}{\int I(\bm{\theta}) \mathrm{d}^2\theta} \; ,
\label{eqn:Q}
\end{equation}
as
\begin{equation}
\left(\begin{array}{c}\epsilon_1 \\ \epsilon_2 \end{array}\right) = \left(Q_{11}+Q_{22}+2\sqrt{Q_{11}Q_{22}-Q_{12}^2}\right)^{-1}
\times\left(\begin{array}{c}Q_{11}-Q_{22} \\ 2Q_{12}\end{array}\right) \; .
\label{eqn:eQ}
\end{equation}
When second moments are measured with uniform weight as in Eqn.~\ref{eqn:Q}, deconvolution with the PSF can be performed by subtracting the observed and PSF second moments,\footnote{This is a simplified procedure for shape measurement, which in practice needs to use appropriate weighting to prevent the divergence of noise. Biases in real shape measurement pipelines should, however, be similar since the primary effect of charge self-interaction is a change in PSF size due to shifting of charges on scales of its FWHM.}
\begin{equation}
Q_{\star}^{\rm gal,\; dec}=Q_{\star}^{\rm gal,\; obs}-Q_{\star}^{\rm PSF} \; .
\label{eqn:QQ}
\end{equation}

Observationally, one estimates $Q_{\star}^{\rm PSF}$ from the second moments of star images. Since the stars used for this are typically at a high flux level in comparison to the galaxy images, the primary effect of charge-self interaction is a misrepresentation of the PSF second moments.\footnote{Note that the effect can be alleviated by using fainter stars only, which are less affected by charge self-interaction (for DECam analyses, cf. e.g. \cite{2014arXiv1405.4285M}).}

In order to simulate the full effect on charge self-interaction on shape measurement, we implement the A14 model in \textsc{GalSim} \cite{2014arXiv1407.7676R}.\footnote{cf. \texttt{https://github.com/GalSim-developers/GalSim/}} Simulating PSF-convolved galaxy and star images with the effect applied, we estimate shapes using Eqns.~\ref{eqn:Q}-\ref{eqn:QQ}. We quantify biases in the measured shape $\epsilon^{\rm meas}$ relative to the input galaxy shape $\epsilon^{\rm true}$ as
\begin{equation}
\label{eqn:bias}
\left(\begin{array}{c}\epsilon_1^{\rm meas}-\epsilon_1^{\rm true} \\ \epsilon_2^{\rm meas}-\epsilon_2^{\rm true}\end{array}\right) =
\left(\begin{array}{c} m_1\epsilon_1^{\rm true} + c_1 \\ m_2\epsilon_2^{\rm true} +c_2 \end{array}\right) 
+ \left(\begin{array}{cc} p^1_1 & p^2_1 \\ p^1_2 & p^2_2 \end{array}\right) \cdot\left(\begin{array}{c} \epsilon^p_1 \\ \epsilon^p_2 \end{array}\right) \; ,
\end{equation}
with multiplicative and additive biases $m_i$ and $c_i$ (e.g. \cite{2006MNRAS.368.1323H}) and leakage terms $p^i_j$ of the PSF ellipticity $\epsilon^p_i$ into the same and respective other component of shear.

For the fiducial simulation settings, we choose the following conservatively realistic parameters. The pre-seeing galaxy and PSF profile are modelled as Gaussians with full-width at half-maximum $\mathrm{FWHM}_{\rm gal}=0.5''$ and $\mathrm{FWHM}_{\rm PSF}=0.9''$, converted to the DECam pixel scale of $0.27''$ per pixel and Gaussian standard deviation $\sigma\approx\mathrm{FWHM}/2.35$. We draw the star image with a $\mathrm{FLUX\_MAX}$ in the central pixel of 60k charges. The galaxy is chosen to have a total flux of 4000 charges. We add no noise to the image, assume a background level of $b=0$ in the fiducial settings and set both galaxy and star image centroid to fall on a pixel center. For the shift coefficients, we use the mean values over the individually fitted chips out to lags $\Delta=8$pix. From determining $\epsilon_{1/2}^{\rm meas}$ at four different settings (where one of the components of either $\epsilon^{\rm true}$ or $\epsilon^p$ is set to $0.1$ and all others are 0) we fix the eight bias parameters of Eqn.~\ref{eqn:bias}. 

Results for the fiducial settings and a number of variants are shown in Table~\ref{tab:bias}. Multiplicative biases are above the per-cent level and exceed the DES weak lensing science requirements. For a hypothetical deep all-sky lensing survey, \cite{2008MNRAS.391..228A} have calculated requirements on systematic errors in shape measurement that translate to approximate limits of $|m|>0.1\times10^{-2}$, $|c|>0.3\times10^{-3}$ and $|p|>0.3\times10^{-2}$, all of which are exceeded by the observed biases.

We probe the seeing range typical for DES lensing band data and object sizes between the smallest usable and median angular size of faint galaxies (cf., e.g., \cite{1999ApJ...527..662P}). We note that multiplicative bias is a strong function of the relative size of PSF and galaxy, as expected due to the PSF size misrepresentation in bright stars. High background levels somewhat alleviate the multiplicative bias since the shifting of background charges is a convolution effect that is independent of the object flux. Interestingly, the strong $x/y$ asymmetry of neighbouring pixel covariances in flat fields (cf. Sec.~\ref{sec:flat}) results in a relatively small additive bias $c_1$ that disappears when we artificially set $a^T_{0,1}$ to the fitted value of $a^R_{1,0}$.

We apply reverse charge shifts to test the appropriateness of our correction scheme, e.g. use shift coefficients $\bm{a}\rightarrow-\bm{a}$ and apply Eqn.~\ref{eqn:Qq}. The purpose of this is to test the influence of a number of simplifications in the corrections we make on shape measurement biases (summarized in the second part of Table~\ref{tab:bias}). Specifically,
\begin{itemize}
\item we correct according to the true model, but only out to lags of $\Delta_{\rm max}=5$pix; multiplicative bias is reduced by a factor of $\approx30$ to $8\times10^{-5}$ for the fiducial case and is, as all other bias parameters, within the requirements;
\item we apply a maximally asymmetric model, degenerate in terms of flat field covariance measurements ($a^T_{0,1}=0$ as transformed with Eqn.~\ref{eqn:degeneracy}), but make a correction for the symmetric model fitted as described in Section~\ref{sec:fitting}; the most significant effect is that additive bias $c_1$ is corrected only partially, but still below science requirements;
\item we apply the model, using a Lanczos-3 interpolated version of the image at the pixel border centers, but for the reverse use the mean counts in the two neighbor pixels in Eqn.~\ref{eqn:Qq}; this leaves a per-mille level multiplicative bias, at the 5\% level of the uncorrected case.
\end{itemize}
We conclude that a $\Delta_{\rm max}=5$, symmetric, linearly interpolated correction scheme is sufficient for our scientific requirements. For a deep all-sky lensing survey, the latter two cause residual biases at the limit of acceptable systematic errors. 

\begin{table*}
\begin{center}
\begin{tabular}{lrrrr}
settings & $m [10^{-2}]$ & $c_1 [10^{-3}]$ & $p^1_1 [10^{-2}]$ & $p^2_2 [10^{-2}]$ \\
\hline \hline
fiducial                        & 2.4 & -0.5 & -0.6 & -0.7 \\ \hline
$\mathrm{FHWM}_{\rm PSF}=0.7''$ & 1.7 & -0.5 & -0.4 & -0.5 \\
$\mathrm{FHWM}_{\rm PSF}=1.1''$ & 3.1 & -0.5 & -0.8 & -0.9 \\ \hline
$\mathrm{FHWM}_{\rm gal}=0.3''$ & 6.9 & -1.4 & -1.7 & -2.0 \\
$\mathrm{FHWM}_{\rm gal}=0.7''$ & 1.1 & -0.3 & -0.3 & -0.3 \\ \hline
$2000$~e$^{-}$ background       & 2.2 & -0.5 & -0.6 & -0.7 \\ \hline
symmetric $a^T_{ij}:=a^R_{ji}$  & 2.5 &  0.0 & -0.6 & -0.7 \\ \hline \hline
corrected out to $\Delta=5$ only& 0.0 &  0.0 &  0.0 &  0.0 \\ \hline
corrected with flat-field degenerate model & 0.0 & -0.2 &  0.0 &  0.0 \\ \hline
corrected neglecting interpolation & -0.1 & 0.0 &  -0.1 &  0.0 \\ 
\hline
\end{tabular}
\caption{Shape biases due to charge self-interaction in DECam as determined from image simulations. Multiplicative biases $m$, additive biases $c$ and PSF leakage $p$, defined as in Eqn.~4.4, are measured with fiducial settings (see text in Section~4), and the described differences in parameters. Note that biases of approximately $|m|>0.1\times10^{-2}$, $|c|>0.3\times10^{-3}$ and $|p|>0.3\times10^{-2}$ are problematic for a hypothetical deep all-sky lensing survey. For DES, requirements are less stringent but clearly exceeded by the observed multiplicative bias. We always find $m=m_1\approx m_2$, $c_2\approx0$ and $p^1_2\approx p^2_1\approx 0$ and omit these from the table.}
\label{tab:bias}
\end{center}
\end{table*}

\section{Correction on pixel-level basis}
\label{sec:correction}

In the previous Sections~\ref{sec:model} and \ref{sec:sim}, we have characterized flat-field noise pixel-to-pixel covariances with the A14 model and shown that the amplitude of charge self-interaction present in DECam is expected to cause shape measurement biases at a critical level. Here, we test whether a correction using our model restores the profiles of bright stars to the shape of their fainter counterparts.

To this end, we analyze a set of 90 r band exposures of moderately dense stellar fields taken between Aug 16, 2013 and Jan 19, 2014. To these we apply overscan and bias subtraction and a correction of inter-CCD cross-talk and signal chain non-linearity. For evaluation of the correction, we generate two versions of the frames, in one of which we apply Eqn.~\ref{eqn:Qq} with the negative of the fitted shift coefficients (e.g. $\bm{a}\rightarrow-\bm{a}$). Finally, we divide by a master flat field.

On the reduced frames we run \textsc{SExtractor} \cite{1996A&AS..117..393B} and split the catalog by brightness. We use the number of counts in the brightest object pixel, \texttt{FLUX\_MAX}, with four bins centered on \texttt{FLUX\_MAX} of 5000, 10000, 15000 and 20000 ADU, all sufficiently below the saturation level of DECam. We run \textsc{PSFEx} \cite{2011ASPC..442..435B} on each of the catalogs to generate a model of the PSF as measured from stars of different flux. We remove a small number of frames for which the fitted PSF models have an irregular appearance visually. We finally stack the PSF models to get the mean PSF profile at each of the flux levels, ensuring that we only use frames where the number of stars in each \texttt{FLUX\_MAX} bin is sufficient to fit a bilinear model of the spatial variation of the PSF that we always evaluate at the chip center position.

\begin{figure}
\centering
\includegraphics[width=\textwidth]{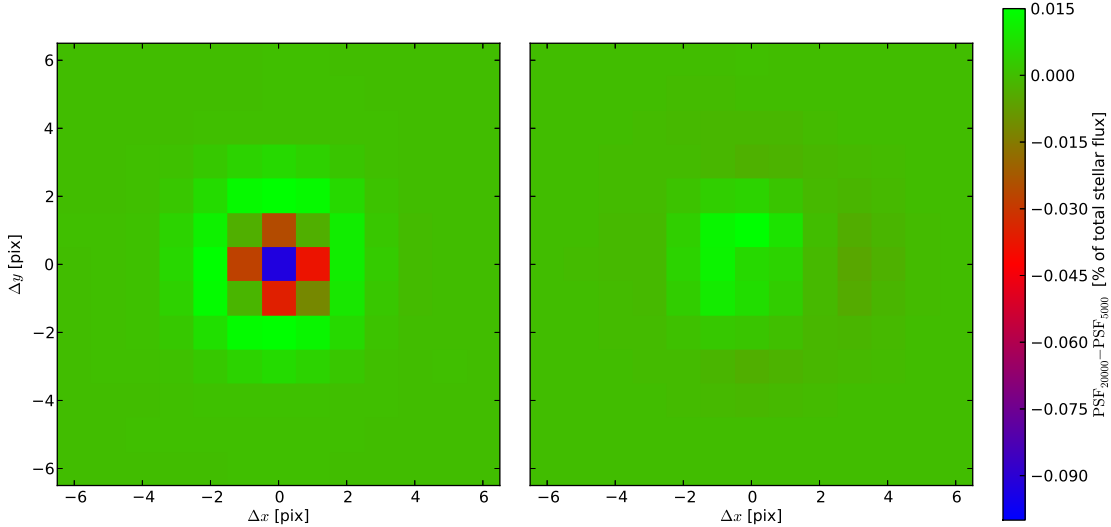}
\caption{Difference image of normalized PSF as measured from stars at \texttt{FLUX\_MAX} of 20000 and 5000 ADU. Left: without correction, the brighter/fatter effect shows as a flux deficit in the central 9 pixels that is compensated by excess charge on an annulus at $\approx$2-4~pix radius. Right: after applying the charge shift correction.}
\label{fig:diffimg}
\end{figure}

Fig.~\ref{fig:diffimg} shows a difference image of the \texttt{FLUX\_MAX}~$\approx5000$ and $20000$ star profile before and after correction. The dominant brighter/fatter feature of a flux deficit in the central pixels is largely removed.

\begin{figure*}
\centering
\includegraphics[width=0.39\textwidth]{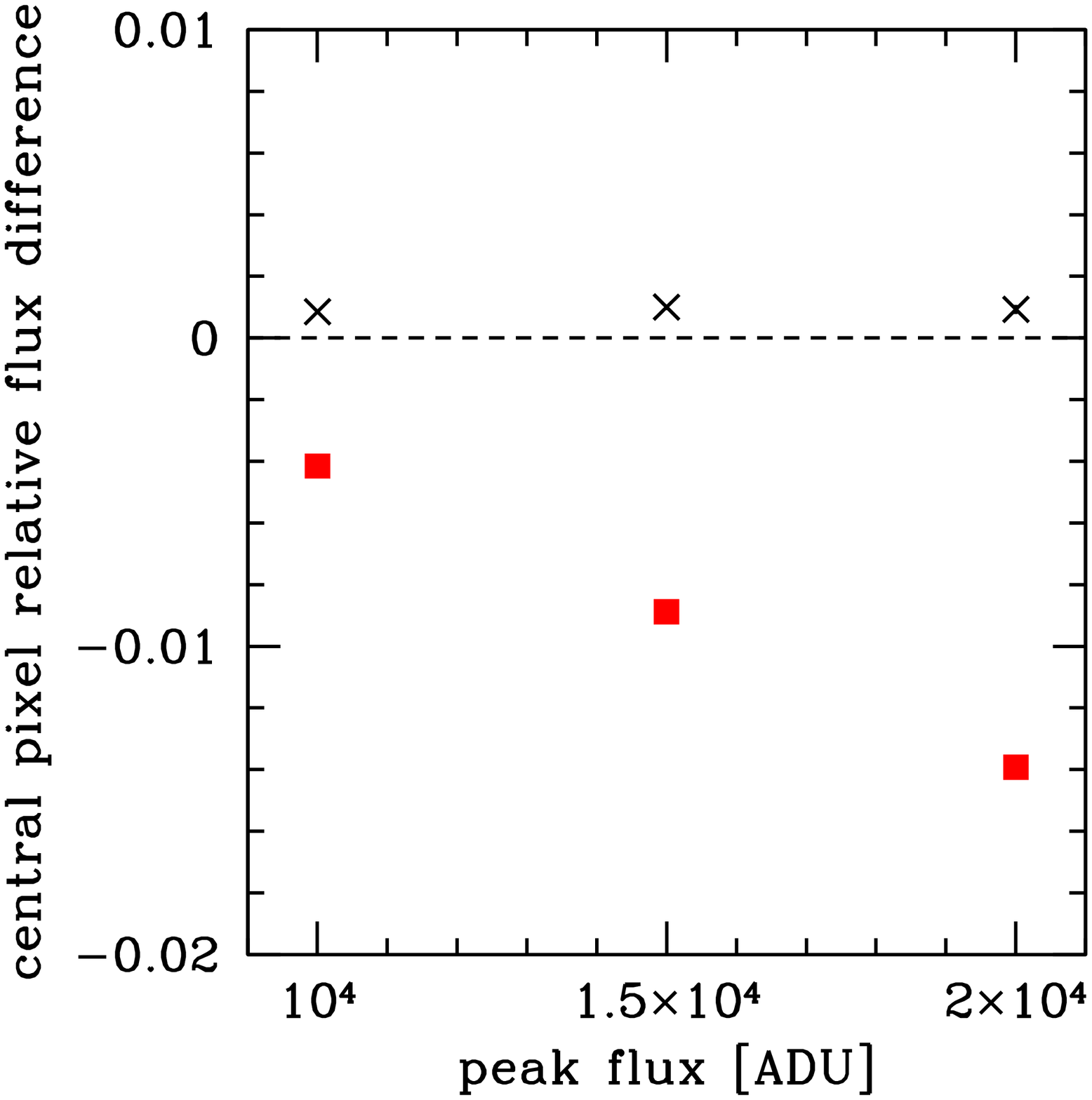}
\includegraphics[width=0.39\textwidth]{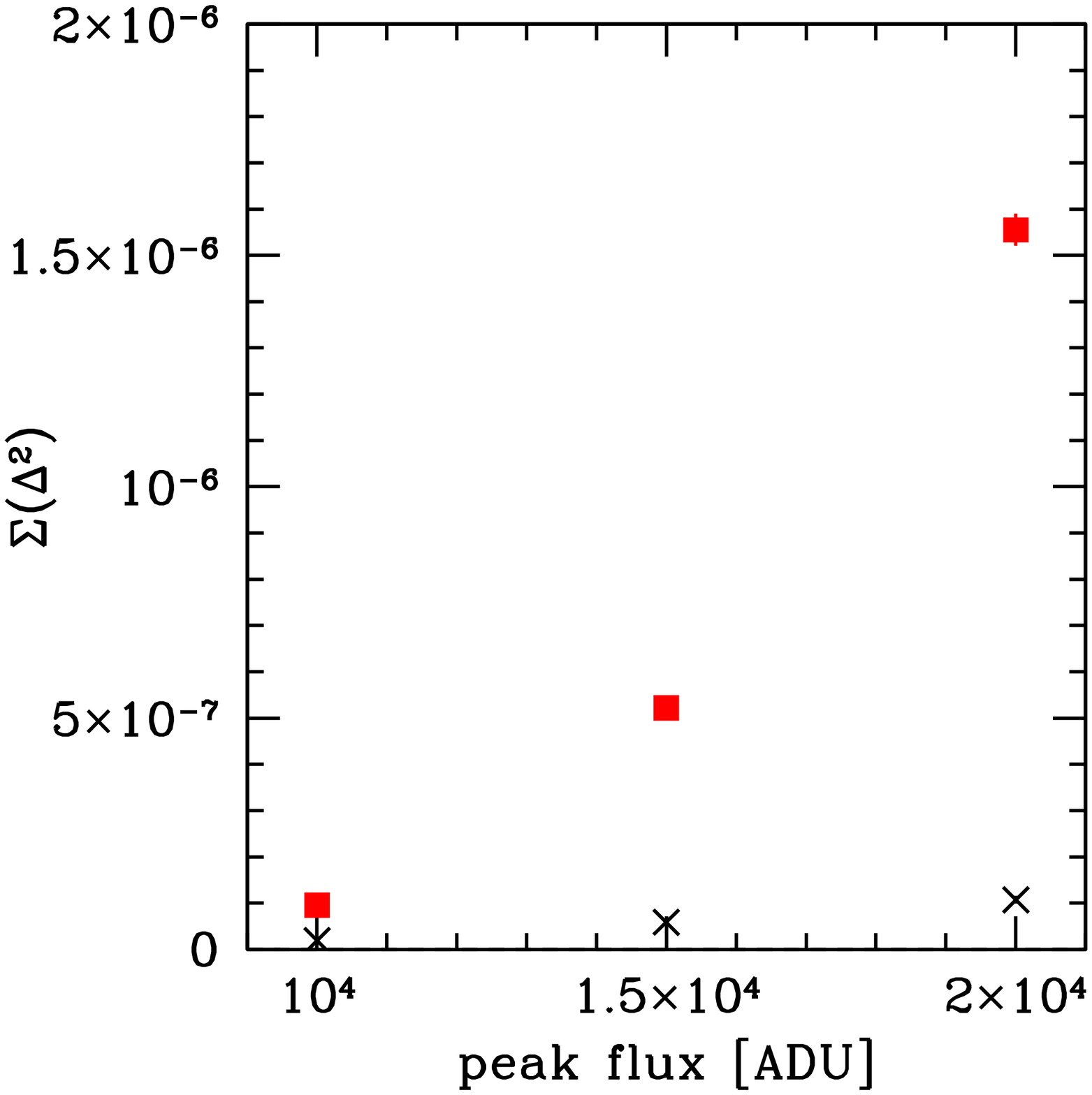}
\caption{Relative change in flux in the central pixel of bright stars (left) and sum of squared residuals between bright and faint stars (right) before correction (red squares) and after applying the reverse charge shifts predicted from the model (black crosses). Both are plotted as a function of peak flux (\texttt{FLUX\_MAX}) and measured relative to stars of \texttt{FLUX\_MAX}~$\approx5$k~ADU with (very small) bootstrapped error bars.}
\label{fig:metrics1}
\end{figure*}

We calculate a number of metrics to quantify how well the correction removes the flux dependence of the PSF. Metrics based on the reduction of the residuals in the individual pixels' fluxes (see Fig.~\ref{fig:metrics1} for flux deficit in central pixel and sum of squared deviations) show a reduction of the effect at the 90 per cent level.

\begin{figure*}
\centering
\includegraphics[width=0.32\textwidth]{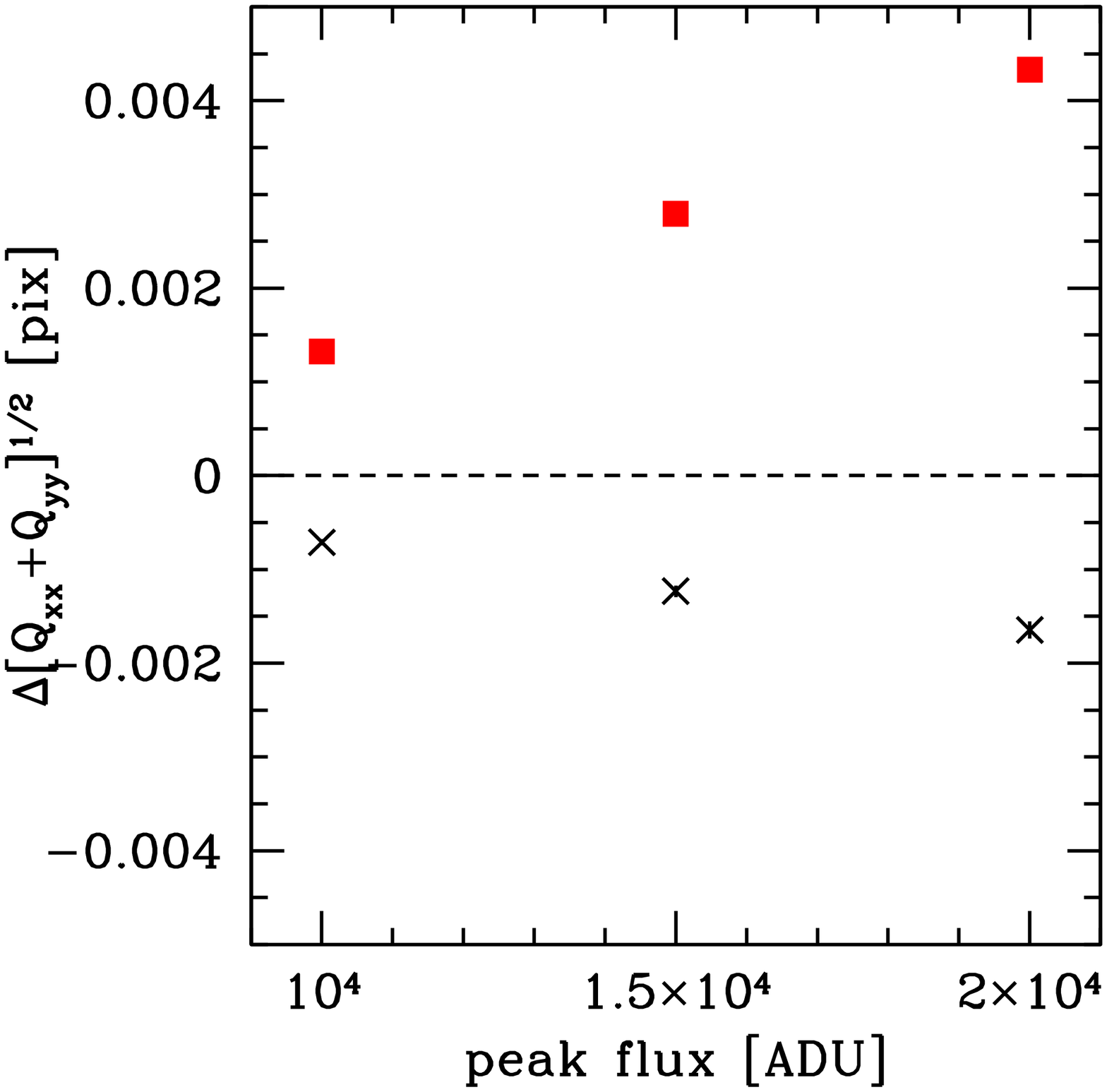}
\includegraphics[width=0.32\textwidth]{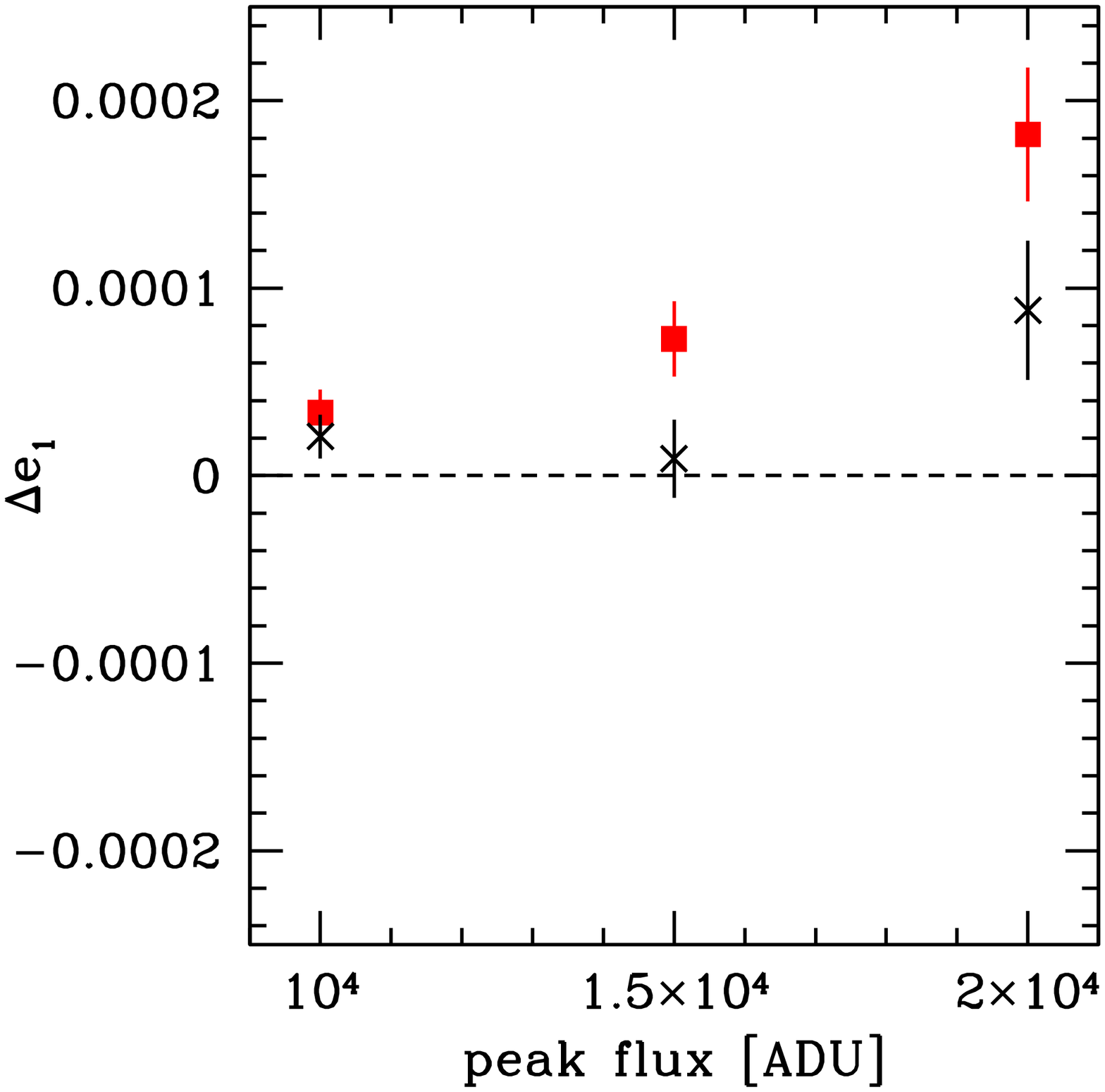}
\includegraphics[width=0.32\textwidth]{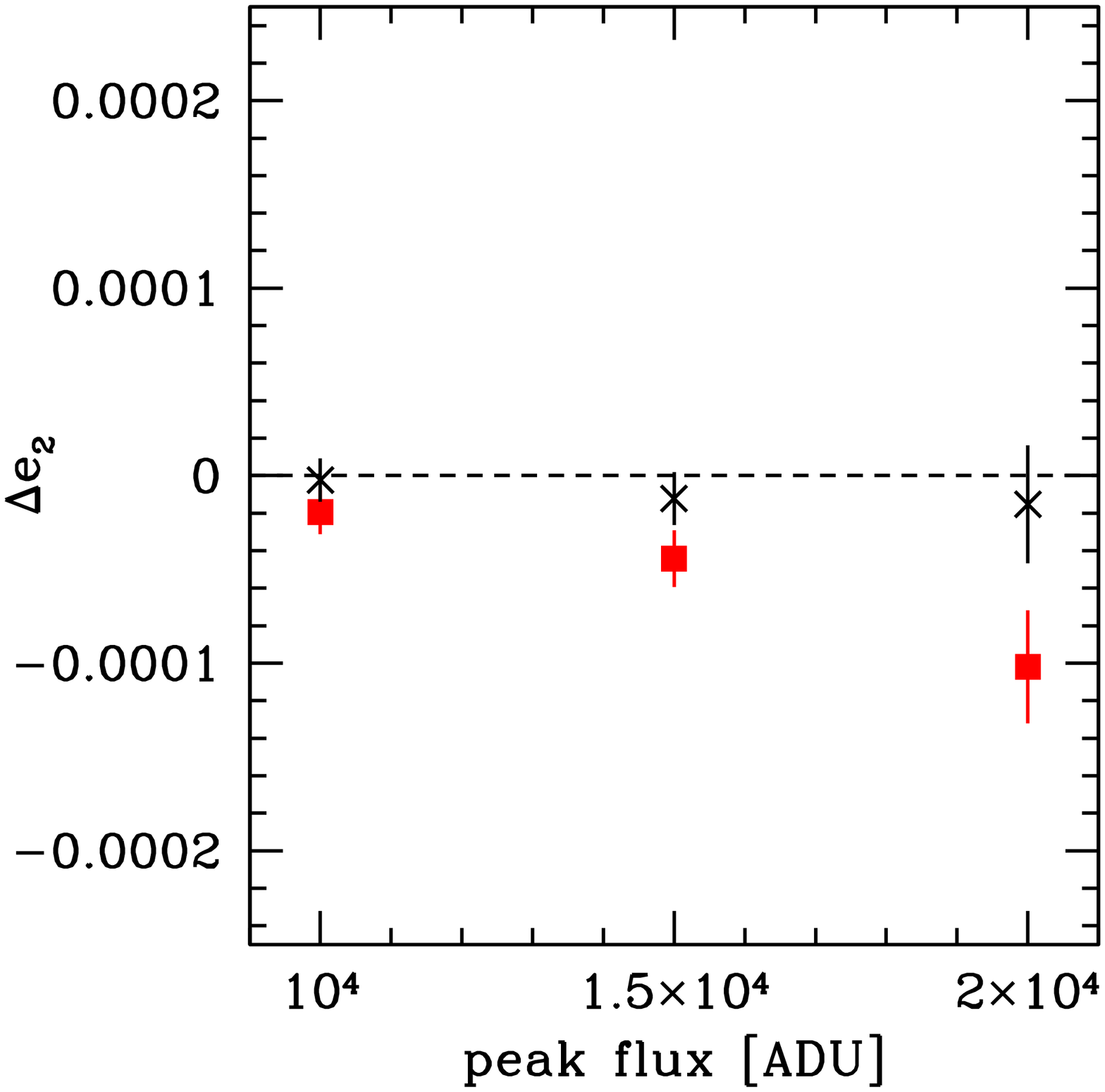}
\caption{Difference in radius (left) and ellipticity (center, right for the two components) between bright and faint stars before correction (red squares) and after applying the reverse charge shifts predicted from the model (black crosses). Quantities are based on second moments $Q_{ij}$ measured with a Gaussian weight function of $\sigma=3~$pix width. Both are plotted as a function of peak flux (\texttt{FLUX\_MAX}) and measured relative to stars of \texttt{FLUX\_MAX}~$\approx5$k~ADU with bootstrapped error bars.}
\label{fig:metrics2}
\end{figure*}

For lensing applications, however, the more relevant quantities are based on second moments of the PSF (cf. Eqns.~\ref{eqn:Q}ff). Figure~\ref{fig:metrics2} shows residuals of three combinations of second moments that correspond to the PSF size and the PSF ellipticity between brighter stars and stars of \texttt{FLUX\_MAX}~$\approx5000$. The residuals at somewhat larger radii that remain in the corrected image (cf. Fig.~\ref{fig:diffimg}) have a significant influence on second moments. While the ellipticity effects of charge self-interaction appear corrected for, size residuals are overcorrected significantly. We have checked that this effect is not due to statistical or systematic differences in centroiding of the PSFEx models. Shape biases associated with the observed residuals after correction are likely still problematic.

We were not able to find the cause of the size residuals at the time of this writing. We hypothesize, however, that they could be connected to either a break-down of the model at low charge levels and large distances or one of the simplifying symmetry assumptions required in constraining the model parameters from flat field covariances (cf. Section~\ref{sec:flat}).

\section{Summary}

We have presented a systematic study of imaging data from DECam and identified two effects of charge self-interaction: (1) a broadening of star images with increasing flux and (2) a correlation of Poisson noise in pairs of pixels in flat field exposures.

The first effect shows as missing flux in the central pixels of bright stars that instead appears at $\approx$3~pix distance from the center. The difference between bright and faint stars is approximately linear in flux and, at fixed total flux level, independent of exposure time (Section~\ref{sec:phenomenology}).

We have measured the second effect with high signal-to-noise ratio out to pixel-to-pixel distances of $\approx$10~pix (Section~\ref{sec:flat}). There is a $\approx5:1$ asymmetry of correlations with the direct neighbor pixels on the same row or column. Correlation amplitudes for more distant pixels are almost isotropic with an approximate power-law fall-off $\propto r^{-2.5}$. Covariances differ from chip to chip at the $\approx20\%$ level. The DECam chips with higher resistivity show larger correlations with their diagonal and distant neighbors and smaller correlations with the adjacent pixel perpendicular to the read-out direction. There is a color dependence of the measured covariances, which is most severe for the latter pairs (more than 40\% increase in correlation from g to Y band) but much weaker for other lags.

Both effects are phenomenologically connected by the A14 charge shift model. From the latter measurements and a number of symmetry assumptions, we fit the model parameters (Section~\ref{sec:fit}). We simulate the effect of the predicted charge shifts on shape measurement (Section~\ref{sec:sim}) and find it to be significant for ongoing Dark Energy Survey weak lensing science, with multiplicative shape measurement biases above the per-cent level.

Finally, we test the ability of the charge shift model to remove the flux dependence of the profiles of stellar images (Section~\ref{sec:correction}). We find that the flux dependence is strongly reduced in terms of the flux deficit in the central pixels and the overall sum of squared residuals. However, significant flux dependence of stellar sizes remains, indicating that our treatment of the charge self-interaction effects on inhomogeneous surface brightness images is yet incomplete.

\acknowledgments

This project was supported by SFB-Transregio 33 `The Dark Universe' by the Deutsche For\-schungs\-gemeinschaft (DFG) and the DFG cluster of excellence `Origin and Structure of the Universe'. DG thanks Pierre Astier, Thomas Diehl, Augustin Guyonnet, Stephen Holland, Mihael Kodric, Ralf Kosyra, and Andy Rasmussen for helpful discussions. GMB acknowledges support for this work from NSF grant AST-1311924 and DOE grant DE-SC007901. AAP is supported by DOE grant DE-AC02-98CH10886 and JPL, which is run under a contract for NASA by Caltech.

We are grateful for the extraordinary contributions of our CTIO colleagues and the DES Camera, Commissioning and Science Verification teams in achieving the excellent instrument and telescope conditions that have made this work possible. The success of this project also relies critically on the expertise and dedication of the DES Data Management organization.

Funding for the DES Projects has been provided by the U.S. Department of Energy, the U.S. National Science Foundation, the Ministry of Science and Education of Spain, 
the Science and Technology Facilities Council of the United Kingdom, the Higher Education Funding Council for England, the National Center for Supercomputing 
Applications at the University of Illinois at Urbana-Champaign, the Kavli Institute of Cosmological Physics at the University of Chicago, Financiadora de Estudos e Projetos, 
Funda{\c c}{\~a}o Carlos Chagas Filho de Amparo {\`a} Pesquisa do Estado do Rio de Janeiro, Conselho Nacional de Desenvolvimento Cient{\'i}fico e Tecnol{\'o}gico and 
the Minist{\'e}rio da Ci{\^e}ncia e Tecnologia, the Deutsche Forschungsgemeinschaft and the Collaborating Institutions in the Dark Energy Survey.

The Collaborating Institutions are Argonne National Laboratory, the University of California at Santa Cruz, the University of Cambridge, Centro de Investigaciones Energeticas, 
Medioambientales y Tecnologicas-Madrid, the University of Chicago, University College London, the DES-Brazil Consortium, the Eidgen{\"o}ssische Technische Hochschule (ETH) Z{\"u}rich, 
Fermi National Accelerator Laboratory, the University of Edinburgh, the University of Illinois at Urbana-Champaign, the Institut de Ciencies de l'Espai (IEEC/CSIC), 
the Institut de Fisica d'Altes Energies, Lawrence Berkeley National Laboratory, the Ludwig-Maximilians Universit{\"a}t and the associated Excellence Cluster Universe, 
the University of Michigan, the National Optical Astronomy Observatory, the University of Nottingham, The Ohio State University, the University of Pennsylvania, the University of Portsmouth, 
SLAC National Accelerator Laboratory, Stanford University, the University of Sussex, and Texas A\&M University.

\end{document}